\documentclass{article}

\usepackage[english]{babel}
\usepackage[dvipsnames]{xcolor}
\usepackage{soul}
\usepackage[letterpaper,top=2cm,bottom=2cm,left=2cm,right=2cm,marginparwidth=1.5cm]{geometry}
\usepackage{caption}
\usepackage{subcaption}
\usepackage{placeins}
\usepackage{amsmath}
\usepackage{amssymb}
\usepackage{graphicx}
\usepackage{multirow}
\usepackage{xcolor}
\usepackage{makecell}
\usepackage{esint}
\usepackage{authblk}
\usepackage{siunitx}
\usepackage[colorlinks=true, allcolors=blue]{hyperref}

\title{On the proper treatment of magnetic fluctuations in full-$f$ field-aligned turbulence codes}
\author[]{Kaiyu Zhang \thanks{corresponding author: kaiyu.zhang@ipp.mpg.de}}
\author[]{Wladimir Zholobenko}
\author[]{Andreas Stegmeir}
\author[]{Konrad Eder}
\author[]{Frank Jenko}
\affil{Max Planck Institute for Plasma Physics, Boltzmannstr. 2, 85748 Garching, Germany}

\date{}
\begin{document}
\maketitle

\begin{abstract}
Plasma turbulence in the edge of magnetic confinement devices is customarily treated as full-\textit{f} due to large fluctuations.
For computational efficiency, field-aligned coordinates are employed, separating the magnetic field into equilibrium $B_0$ and delta-\textit{f} perturbations which are handled by the magnetic flutter operators.
Evolving the full-\textit{f} pressure with delta-\textit{f} magnetic perturbations can cause inconsistency since the latter contain background components such as the Shafranov shift, which are actually parts of the equilibrium magnetic field. 
Such background components ($B_s$) contained in the magnetic perturbations undermine the field-aligned numerics when treated as flutter: errors arise if $B_s/B_0\ll l_\perp/h_\parallel$ is not satisfied, with the perpendicular turbulence scale $l_\perp$ and the parallel grid distance $h_\parallel$. 
We find that the commonly used removal of $B_s$ by subtracting the toroidal average of magnetic perturbations intervenes in the Alfv\'en dynamics, causing spurious $E\times B$ transport.
Instead, we propose an improved method to dynamically filter out the evolving background from the turbulent magnetic fluctuations in the time domain. 
The filter is verified in both low and high confinement tokamak conditions, confirming its capability to preserve the turbulence fidelity, provided sufficient filter width.
\end{abstract}

\section{Introduction} \label{sec:intro}
Plasma turbulence models generally fall into two categories: `delta-\textit{f}' if the fluctuation part $\delta f$ is separated from the background $f_0$, and `full-\textit{f}' otherwise \cite{scott2021turbulence}. 
$f$ here is not necessarily the particle distribution function, but any quantity. 
The former models are computationally simpler and particularly suited for the core region of tokamaks and stellarators. On the other hand, in the edge and particularly in the scrape-off layer (SOL), the fluctuation amplitudes of density and temperature are large \cite{Zweben2007, Boedo2009, zholobenko2023filamentary}, $\delta f /f_0 \sim 1$, the plasma and the magnetic geometry are spatially very inhomogeneous and the transport time scales (profile evolution) are much closer to those of the turbulence saturation. 
Consequently, most codes for edge-SOL turbulence simulations adopt a full-\textit{f} approach for the plasma \cite{stegmeir2019global, giacomin2022gbs, zhu2018gdb, bufferand2021progress, Dudson2024, hager2022electromagnetic, michels2022full, mandell2022turbulent}. 
Magnetic fluctuations are known to be crucial in edge turbulence \cite{Dickinson2012,scott2021turbulence,zholobenko2024tokamak}. 
However, they are commonly treated separately under the delta-\textit{f} approximation for two primary reasons. 
First, the equilibrium magnetic field in tokamaks is typically 3 to 4 orders of magnitude \cite{hidalgo1995edge, zhu2023electromagnetic, zhang2024magnetic, zholobenko2024tokamak} greater than the magnetic fluctuations induced by small-scale turbulent plasma currents. 
Second, to enhance computational efficiency, many codes align their coordinates and operators with the constant equilibrium magnetic field to keep the coordinate along the magnetic field sparse, thereby permitting larger time steps \cite{STEGMEIR2023108801,Michels2021}.

In this setting, parallel operators are constructed on the grid based on the equilibrium magnetic field $\mathbf{B}_0$. The flutter operators \cite{zhang2024magnetic} then account for magnetic fluctuations $\mathbf{B}_1$ within the turbulence. 
However, coupled with the full-\textit{f} treatment for the plasma, this setup can introduce a subtle inconsistency. 
This is detailed in section \ref{sec:reason}, but in short, full-$f$ models can contain parts of the equilibrium current, such as the Pfirsch-Schlüter current. Such currents lead to a large-scale shift of magnetic surfaces known as the Shafranov shift \cite{helander2005collisional}, which is known to be important both for stability \cite{Snyder2007} and transport \cite{Lackner2000}. More generally, full-\textit{f} (gyro-)kinetic models also contain the bootstrap current \cite{Hager2016}, while in the SOL, thermal currents generated by the stationary temperature gradient are expected \cite{kallenbach2001electric}. These are all macroscopic currents with near zero frequency, compared to the higher-frequency turbulence dynamics. 
As a result, the magnetic field induced by these currents does not represent a true delta-\textit{f} quantity. Instead, the magnetic perturbations carry a near-zero frequency offset, referred to in this paper as the ``magnetic shift". 
The primary issue lies in the constant redirection of the background magnetic field due to the magnetic shift, which ultimately reconstructs the field-aligned structure. 
By definition, the magnetic shift should have been included in the equilibrium magnetic field $\mathbf{B}_0$.
Without careful differentiation between $\mathbf{B}_0$ and $\mathbf{B}_1$, the magnetic shift would be double-counted in the parallel operators and flutter terms.

This inconsistency has been widely acknowledged across the drift-fluid \cite{giacomin2022gbs, dull2024introducing}, gyro-fluid \cite{scott2006edge} and gyro-kinetic \cite{hager2022electromagnetic} turbulence communities, though no unified solution has been found yet. The issue of double-counting the magnetic shift can be addressed in two ways: either by removing the magnetic shift from the equilibrium field or by subtracting it from the flutter operators.
In the first solution, magnetic flutter exclusively incorporates the magnetic shift.
$\mathbf{B}_1$ is then directly solved from the parallel magnetic potential $A_\parallel$, enduring its deviation from delta-\textit{f}.
This solution could be feasible for non-aligned codes such as GBS \cite{giacomin2022gbs}, provided the toroidal resolution is sufficient.
However, for field-aligned codes, the amplitude of magnetic perturbations is a critical factor when treated by flutter terms.
In section \ref{sec:limit}, we present a theoretical framework to predict the numerical constraints on a large magnetic shift in field-aligned codes. Specifically, we show that incorporating the magnetic shift into flutter terms becomes impractical under high-beta conditions.

The second approach is to remove the magnetic shift from the flutter operators, 
maintaining $\mathbf{B}_1$ as purely a delta-$f$ fluctuation. The challenge arises in dynamically defining and adjusting the magnetic shift during runtime.
A key characteristic of full-\textit{f} turbulence simulations is the temporal evolution of the background plasma profiles, which gradually deviate from their initial assumptions. 
As a result, the magnetic shift, being intrinsically linked to the background plasma profiles, cannot remain stationary but must evolve correspondingly.
The Pfirsch-Schlüter structure in the magnetic shift can be analytically determined using the ballooning angle representation \cite{scott2006edge,ribeiro2007self}.
However, this approach is less reliable near the X-points. In realistic geometries that include the X-point, comprehensive models with multiple background $A_\parallel$ components often adopt a posterior method of taking the instantaneous toroidal average of $A_\parallel$ as the background \cite{giacomin2022gbs,dull2024introducing,hager2022electromagnetic}.
Hager \textit{et al.} \cite{hager2020gyrokinetic} raised concerns that an instantaneous toroidal average might retain residual turbulent structures.
While the toroidal average method has been employed in many simulations, its interactions with turbulent dynamics and the extent to which it may compromise physical fidelity have not been rigorously examined.
In section \ref{sec:sim}, we present the first analysis demonstrating that this method of instantaneous toroidal averaging indeed introduces appreciable numerical artifacts, indicated by a significant increase in turbulence transport in the L-mode ASDEX Upgrade (AUG) tokamak edge, with even more profound and complex effects in H-mode cases.

In this paper, we first illustrate in section \ref{sec:reason} the overlap between the magnetic equilibrium and the full-\textit{f} electromagnetic turbulence which leads to the emergence of the magnetic shift. 
Section \ref{sec:limit} compares two different treatments of the magnetic shift, using the magnetic flutter operator and tracing the perturbed field line, within GRILLIX simulations. 
The limitation of the flutter operator is formalized in Eqn.~\eqref{eqn:flutter_limit}, providing a criterion for determining whether the magnetic shift should be incorporated into the equilibrium field or treated through flutter terms.
The limitation has been tested in the examples of GRILLIX, but the ideas can simply be transferred to any locally or globally field aligned description.
Section \ref{sec:method} introduces a new method to filter out the magnetic shift, isolating the pure delta-\textit{f} magnetic fluctuations within the flutter terms, thereby achieving self-consistency in full-\textit{f} turbulence simulations. 
This implementation is straightforward and highly portable. The filter dynamically tracks and removes the evolving background $A_\parallel$ in real-time during simulations, consistently mitigating strong magnetic shifts within flutter terms and avoiding the associated numerical issues.
In Section \ref{sec:sim}, the influence of this filter on microinstabilities and turbulent transport is analyzed through simulations for the edge and scrape-off layer of AUG in both L-mode and H-mode conditions.
We conclude that sufficient filter widths, formalized as Eqn.~\eqref{eqn:lambda_limit}, can effectively preserve the turbulence fidelity, outperforming the traditional method of toroidal average.

\section{Magnetic shift in full-$f$ models}\label{sec:reason}
The magnetic field $\mathbf{B}_0$, current $\mathbf{j}_0$ and plasma pressure $p_0$ in equilibrium are governed by the force balance equation
\begin{equation} \label{eqn:equi:balance}
    \mathbf{j}_0\times \mathbf{B}_0 = \nabla p_0\,.
\end{equation}
In a tokamak, this leads to the Grad-Shafranov equation \cite{zohm2014magnetohydrodynamic}.
The perpendicular current can be expressed as 
\begin{equation} 
    \mathbf{j}_{0,\perp}  = \mathbf{B}_0 \times \nabla p_0 / |B_0|^2 \,.
    \label{eq:j_dia}
\end{equation}
Due to the curvature of the toroidal magnetic field, this diamagnetic current has a non-zero divergence $\nabla \cdot \mathbf{j}_{0,\perp} = -C(p_0)$, where $-C(p_0) = \nabla\cdot \left( \mathbf{B}_0 \times \nabla p_0 / B_0^2 \right) = \left( \nabla \times \mathbf{B}_0/B_0^2 \right) \cdot \nabla p_0$ is the curvature operator \cite{zholobenko2021electric}. 
To satisfy charge conservation and quasi-neutrality, a parallel current emerges
\begin{equation} \label{eqn:equi:ps_current}
    \nabla \cdot j_{0,\parallel} \mathbf{b}_{0} = C(p_0) \,.
\end{equation}
This is the so-called Pfirsch-Schlüter (PS) current \cite{hirshman1978neoclassical}.
In the closed field line region of a tokamak, the equation \eqref{eqn:equi:ps_current} can be solved analytically \cite{hirshman1978neoclassical}
\begin{equation} \label{eqn:equi:ps_current_ana}
    j_{0,\parallel}= -\frac{2\pi}{B_0}\frac{\mathrm{d}p_0}{\mathrm{d}\psi} \left(1-\frac{B_0^2}{\langle B_0^2\rangle_\theta}\right) \,,
\end{equation}
in which $\langle \rangle_\theta$ means the flux surface average, and $p_0$ in equilibrium is a function of the poloidal flux $\psi$.
Through Ampère’s law, it induces a magnetic field that shifts the magnetic flux surfaces radially outward, forming a part of the Shafranov shift\cite{sato2000complete} (the poloidal field coils provide the other part \cite{zohm2014magnetohydrodynamic}).

\begin{figure}[!hbp]
\centering
\includegraphics[width=0.6\linewidth]{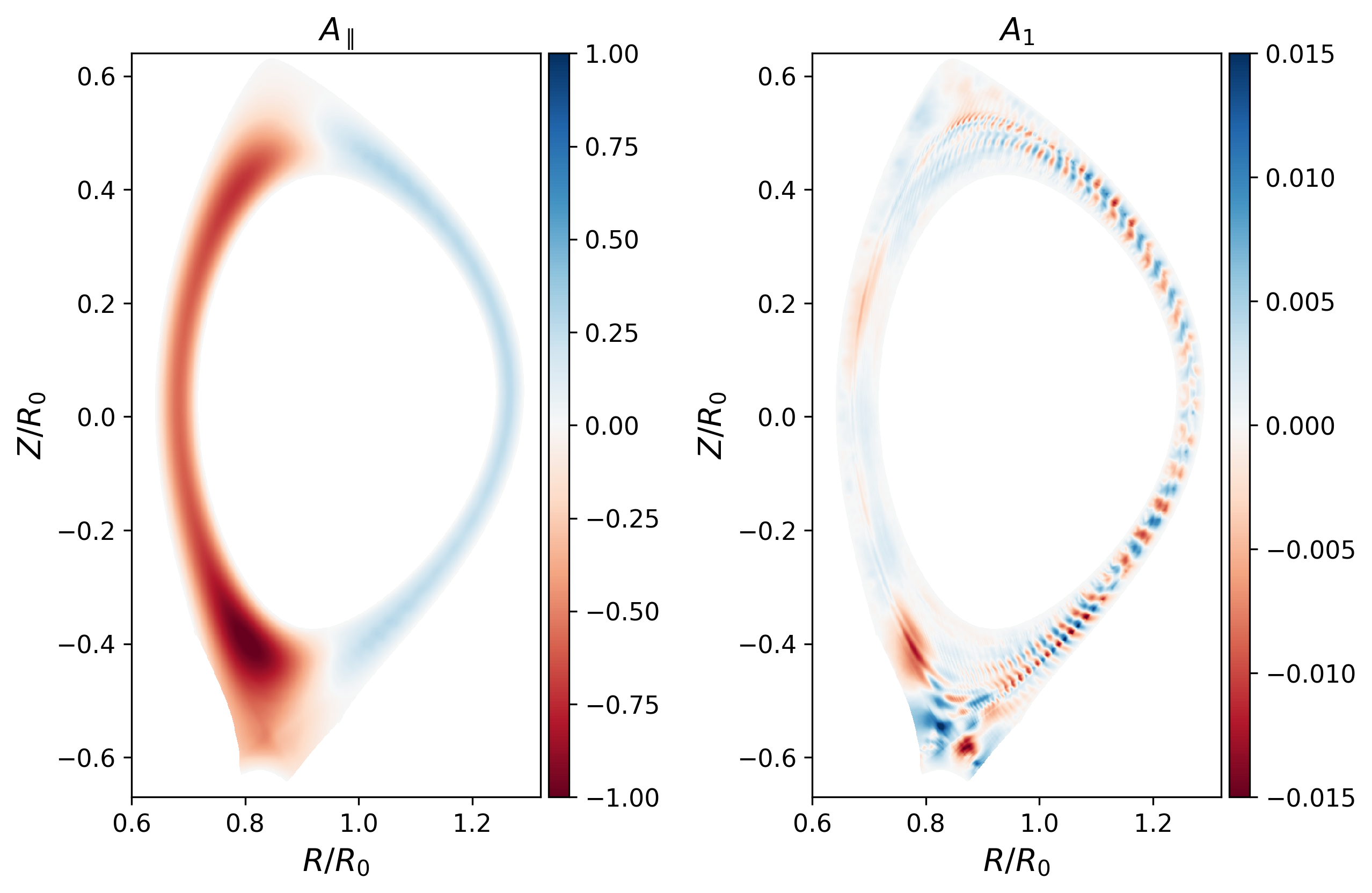}
\caption{Instantaneous snapshots of $A_\parallel$ (left) and its fluctuation $A_1$ (right) in H-mode conditions \cite{zholobenko2024tokamak}, normalised to $A_\mathrm{ref}=\rho_{s0} B_\mathrm{axis}=1.44\times10^{-3}\mathrm{m}\cdot\mathrm{T}$. $A_1$ is obtained using the filter Eqn.~\eqref{eqn:evol_apar} and \eqref{eqn:evol_A_1} introduced in section \ref{sec:method}. }
\label{fig:apar_2D_comp}
\end{figure}

GRILLIX employs the 6-field global full-$f$ (extended) Braginskii equation set (its most recent version is documented in \cite{zholobenko2024tokamak}) to describe the plasma dynamics, including turbulence.
The quasi-neutrality equation reads
\begin{equation} \label{eqn:vort}
    \nabla \cdot j_{\parallel} \mathbf{b} = C(p) + \nabla \cdot n \mathbf{u}_\mathrm{p} \, ,
\end{equation}
where $\mathbf{u}_\mathrm{p}$ is the ion polarisation velocity, responsible for the turbulence \cite{zholobenko2021electric}.  $n$ is the plasma density and
$\mathbf{b}$ is the unit vector of the magnetic field including fluctuations. 
$j_\parallel$ is coupled to the parallel magnetic potential $A_\parallel$ by Ampère’s law
\begin{equation} \label{eqn:amper}
    \nabla^2_\perp A_{\parallel} = -j_\parallel \, .
\end{equation}
$A_\parallel$ and $j_\parallel$ are solved via Ohm's law
\begin{equation} \label{eqn:ohm}
    \beta_0 \frac{\partial}{\partial t} A_{\|}+\mu\left(\frac{\partial}{\partial t}+\mathbf{v}_{E}+v_{\|} { \mathbf{b}\cdot\nabla}\right) \frac{j_{\|}}{n}
    =-\left(\frac{\eta_{\| 0}}{T_{e}^{3 / 2}}\right) j_{\|}-{\mathbf{b}\cdot\nabla} \phi+\frac{\mathbf{b}\cdot\nabla p_{e}}{n}+0.71 {\mathbf{b}\cdot\nabla}T_e\,,
\end{equation}
where $\mu$ is the mass ratio of the electron to ion, $\mathbf{v}_{E}$ is the $E\times B$ velocity, $v_{\|}$ is the electron parallel velocity, and $\eta_{\| 0}$ is the dimensionless parallel resistivity. $\phi$ is the electrostatic potential. $p_e$ and $T_e$ are the electron pressure and temperature, respectively.
Importantly, our full-$f$ model evolves the total pressure including both the background $p_0$ and fluctuations $p_1$, namely $p=p_0+p_1$. Therefore, clearly, through $C(p)$ in \eqref{eqn:vort} and $j_\parallel$ in \eqref{eqn:amper}, the Shafranov shift is also contained in $A_\parallel$ in our model. 
The Pfirsch-Schlüter current is important in global turbulence simulations for various reasons, including global geodesic Alfvén modes \cite{scott2008fully} and the influence of the PS current on the edge electric fields \cite{aydemir2012pfirsch}.
However, this adds the complication of double-counting the Shafranov shift in parallel operators, if it is contained in both the magnetic equilibrium and our dynamically evolved $A_\parallel$.
Fig.~\ref{fig:apar_2D_comp} (left) shows a 2D snapshot of $A_{\parallel}$ from an H-mode simulation of the ASDEX Upgrade (AUG) tokamak \cite{zholobenko2024tokamak}.
The Shafranov shift with a poloidal mode number $m=1$ is evident.
The actual turbulent fluctuations $A_1$ on top of the Shafranov shift, shown in Fig.~\ref{fig:apar_2D_comp} (right), are nearly 2 orders of magnitude smaller than $A_{\parallel}$. 
In this paper, the parallel magnetic potential is consistently normalized to $A_\mathrm{ref} =\rho_{s0} B_\mathrm{axis}=1.44\times 10^{-3} \mathrm{m}\cdot\mathrm{T}$.
$B_\mathrm{axis}$ is the magnetic field at the magnetic axis.
$\rho_{s0}=\sqrt{m_\mathrm{i}T_\mathrm{ref}}/(e B_\mathrm{axis})$ is the nominal sound Larmor radius, where $m_\mathrm{i}$ is the ion mass and the normalisation for the electron temperature is $T_\mathrm{ref}=100\mathrm{eV}$.

From this perspective, we define the \textit{magnetic shift} $A_s$ (with $\mathbf{B}_s = \nabla \times A_s \mathbf{b}_0$) as the parallel magnetic potential (and the corresponding perpendicular magnetic field) associated with the background plasma profiles within the full-\textit{f} model.
This definition encompasses not only the PS-current-driven Shafranov shift but also the magnetic field contributions arising from the $\nabla_\parallel T_{e,0}$ driven thermal current in the SOL, for example. Furthermore, in a full-$f$ kinetic model, the magnetic shift may include components induced by the bootstrap current.
Given its dependence on multiple physical processes, $A_s$ cannot be easily expressed analytically as a function of the background profiles $n_0$ or $T_0$, particularly in the complex geometries present near the X-point and separatrix. Moreover, $A_s$ is not static; instead, it evolves slowly in tandem with the changes in the background plasma profiles. To describe this evolution, we define $\omega_s$ as the characteristic frequency of background profile evolution.
The frequency $\omega_s$ must be much smaller than the turbulent frequency $\omega_t$, which typically approaches the Alfvén frequency $\omega_A$:
\begin{equation}
     \omega_s \ll \omega_t \sim \omega_A
\end{equation}
This low-frequency evolution of $A_s$ introduces both physical inconsistencies and potential challenges for field-aligned codes, as will be demonstrated in the subsequent section.

\section{Numerical limitations on magnetic flutter in field-aligned codes}\label{sec:limit}

In the presence of a strong background magnetic field, plasma turbulence has predominantly the flute mode character $k_\parallel/k_\perp \ll 1$, where $k_\parallel$ and $k_\perp$ denote the parallel and perpendicular wave numbers, respectively) \cite{scott2021turbulence}.
Field-aligned coordinates \cite{stegmeir2016field} are able to significantly reduce the parallel grid resolution of $\nabla_{\parallel}$
and relax the CFL condition restricted by the shear Alfv\'en dynamics, yielding a computational speed-up which scales as $(k_\perp/k_\parallel)^2$ \cite{STEGMEIR2023108801} compared to the non-aligned method.
A strong guide magnetic field is required in the first place to which the coordinates are aligned, which is typically provided by the magnetic equilibrium $\mathbf{B}_0$ (the unit vector is $\mathbf{b}_0$) in fusion devices.
The parallel operator in equilibrium $\nabla_\parallel^\mathrm{equi}=\mathbf{b}_0\cdot\nabla$ is constructed via tracing field lines \cite{stegmeir2016field}.
This raises a critical question: what happens in the case of  a magnetic perturbation?

\begin{figure}[!htbp]
\centering
\includegraphics[width=0.35\linewidth]{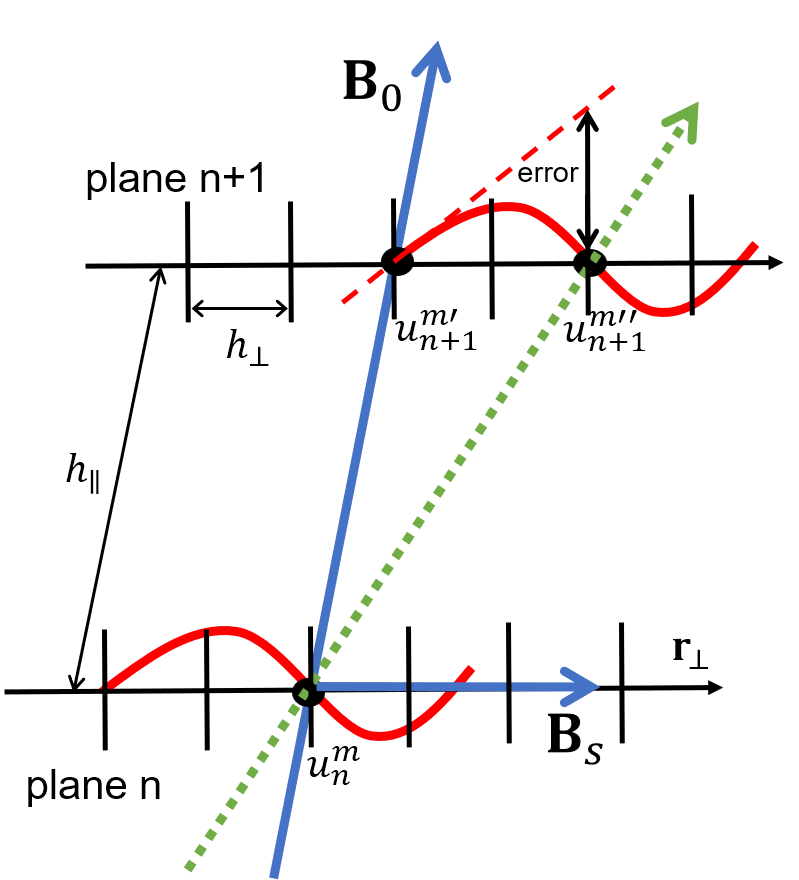}
\caption{\label{fig:demo_flutter_approx} Illustration of the magnetic perturbations in the (locally) field-aligned framework. The black lines represent two neighboring poloidal planes and the grid stencil has an arrow pointing in a perpendicular direction. 
The blue arrows are the background magnetic field $\mathbf{B}_0$ and the magnetic shift $\mathbf{B}_s$ as the perturbation.
The original field alignment $u_n^m \rightarrow u_{n+1}^{m'}$ is established by $\mathbf{B}_0$.
The green dotted line shows the new alignment $u_n^m \rightarrow u_{n+1}^{m''}$ redirected by $\mathbf{B}_s$.}
\end{figure}

The magnetic perturbations occurring in our simulations fall into two types, contingent upon whether they reshape the field alignment:
\begin{enumerate}
    \item Turbulent magnetic fluctuation, denoted $\mathbf{B}_1$, have the typical frequencies $\omega_t \sim \omega_A$. As a pure fluctuation, $\mathbf{B}_1$ can not effectively change the field alignment that has been still established by $\mathbf{B}_0$. 
    Instead, $\mathbf{B}_1$ appears as a perpendicular drift effect and contributes to the quasi-static force balance across the background field lines \cite{scott2021turbulence}. Therefore, the associated derivatives $\mathbf{B}_1\cdot \nabla$, namely magnetic flutter, depend on the local gradient, which is discretised within individual perpendicular planes.
    The appendix \ref{appe} documents in detail the implementation of the magnetic flutter in GRILLIX, ensuring $\nabla \cdot \mathbf{B}_1 = 0$.
    \item The magnetic shift $\mathbf{B}_s$, defined in section \ref{sec:reason}, has a slow frequency $\omega_s \ll \omega_A$. This case is shown in Fig.~\ref{fig:demo_flutter_approx}. The magnetic shift will fundamentally rebuild the alignment of parallel dynamics $u_n^m \rightarrow u_{n+1}^{m''}$ along the direction $\mathbf{B}_0+\mathbf{B}_s$.
    This effect has been observed in our previous simulation for the stellarator in Fig.~9 of \cite{stegmeir2024stell}. 
    Isolating $\mathbf{B}_s \cdot \nabla$ within individual planes as the magnetic flutter terms becomes questionable, of which the limitation will be critically analysed in the following.
\end{enumerate}

Now we focus on the magnetic shift case. 
Fig.~\ref{fig:demo_flutter_approx} illustrates the effect of magnetic shift in the locally field-aligned system. 
The parallel grid distance between two points is $h_\parallel$, and the perpendicular grid distance is $h_\perp$.
The plasma field $u$ (for instance, temperatures or density) is regarded as turbulent, of which fluctuations are represented by the red curve with a perpendicular scale of approximately $l_\perp$.
The equilibrium field line $\mathbf{B}_0$ connects a point $u_{n}^{m}$ on the $n$-th plane and the mapping point $u_{n+1}^{m'}$ on the $n+1$ plane.
On top of the low-frequency nature of $\omega_s \ll \omega_A$,
we make the further assumption that $\mathbf{B}_s$ is axisymmetric in the toroidal direction, with a larger perpendicular scale $l_{s}\gg l_\perp$.
This assumption holds for the Shafranov shift in equilibrium.
After being perturbed by $\mathbf{B}_s$, the magnetic field line redirects towards the point $u_{n+1}^{m''}$.
In the view of finite difference schemes, the full parallel gradient is 
\begin{equation}\label{eqn:flutter_exact}
    \nabla_\parallel u = \underbrace{\frac{u_{n+1}^{m''}-u_{n}^{m}}{h_\parallel}}_{\nabla_{\parallel}^\mathrm{retrace}}=\underbrace{\frac{u_{n+1}^{m'}-u_{n}^{m}}{h_\parallel}}_{\nabla_{\parallel}^\mathrm{equi}}+\underbrace{\frac{u_{n+1}^{m''}-u_{n+1}^{m'}}{h_\parallel}}_{\frac{\mathbf{B}_s\cdot\nabla}{B_0}}.
\end{equation}
We ignore the change of $h_\parallel$ caused by $\mathbf{B}_s$, because $B_s \ll B_0$. 
The displacement of $u_{n+1}^{m''}$ with respect to $u_{n+1}^{m'}$ is $\mathbf{r}_s = h_\parallel \mathbf{B}_s/B_0$.
We try to approach $u_{n+1}^{m''}$ utilizing the local information around the point $u_{n+1}^{m'}$.
The Taylor expansion along the direction of $\mathbf{r}_s$ is 
\begin{equation}
    u_{n+1}^{m''} = u_{n+1}^{m'}+ \frac{\partial u_{n+1}^{m'}}{\partial r_\perp} r_s+O({r}_s ^2)
\end{equation}
hints at the first-order approximation
\begin{equation}\label{eqn:flutter_approx}
    \underbrace{\frac{u_{n+1}^{m''}-u_{n+1}^{m'}}{h_\parallel}}_{\nabla_{\parallel}^\mathrm{retrace}-\nabla_{\parallel}^\mathrm{equi}} \approx \underbrace{\frac{\partial u_{n+1}^{m'}}{\partial r}\cdot\frac{B_s}{B_0}}_{{\nabla^\mathrm{flutter}}}
\end{equation}
The right-hand side is exactly the flutter term applied on the point $u_{n+1}^{m'}$ (the full expression \ref{eqn:flutter_grad} in Appendix \ref{appe}), which follows the red dashed line in Fig.~\ref{fig:demo_flutter_approx} as $B_s$ varies.
The flutter terms are linear in the magnitude of the magnetic perturbation (they are nonlinear in the sense that both $u$ and $B_s$ are perturbed quantities).
The approximation \eqref{eqn:flutter_approx} will break and the error will surge when $r_s$ becomes comparable to the scale of turbulence $l_\perp$, as depicted by the gap between the red dashed line and curve in Fig.~\ref{fig:demo_flutter_approx}.
Importantly, this limitation can be formulated as the quantity $\kappa$ being small,
\begin{equation}\label{eqn:flutter_limit} 
    \kappa=\frac{B_s}{B_0}\frac{h_\parallel}{l_\perp}\ll 1.
\end{equation}
There are two important points to be noted: 
(1) The limitation \eqref{eqn:flutter_limit} depends on the parallel grid distance $h_\parallel$ instead of the parallel length of the turbulence $l_\parallel$.
In principle, one can always reduce $\kappa$ by increasing the toroidal resolution, for the price of perhaps losing some of the advantages of field alignment.  
(2) $l_\perp$ is subject to the physical micro-instability, for instance, $l_\perp<\rho_s$ in the electron temperature gradient (ETG) driven turbulence and $l_\perp>\rho_s$ in the ion temperature gradient (ITG) turbulence.
However, it must be bounded by the resolution, namely $l_\perp \ge h_\perp$.
With a large $h_\perp$ the small scale mode may not be resolved, but the flutter approximation may still be numerically valid.
In practice, it is difficult to know in advance the accurate $l_\perp$. In global simulations, $l_\perp$ is also varying in space (typically smaller on the low field side but larger on the high field side). Therefore it is useful to define a nominal $\hat{\kappa}$, which is the upper bound of $\kappa$ limited by the perpendicular resolution:
\begin{equation}\label{eqn:flutter_limit_nom} 
    \hat{\kappa}=\frac{B_s}{B_0}\frac{h_\parallel}{h_\perp} \ll 1
\end{equation}

\begin{figure}[!htbp]
\centering
\begin{subfigure}[b]{0.4\textwidth}
    \includegraphics[width=\textwidth]{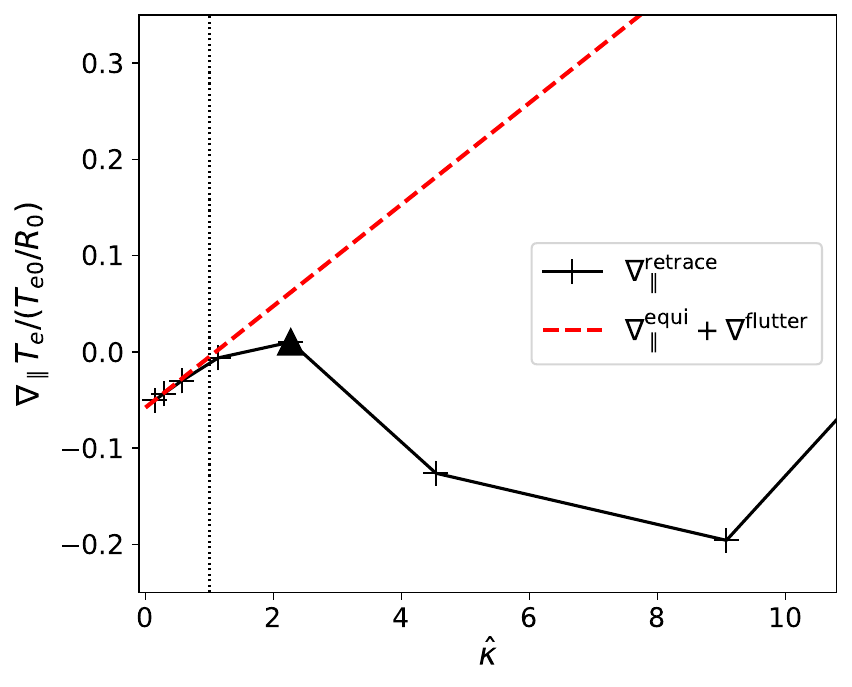}
    \caption{\label{fig:flutter_approx_single}}
\end{subfigure}
\begin{subfigure}[b]{0.4\textwidth}
    \includegraphics[width=\textwidth]{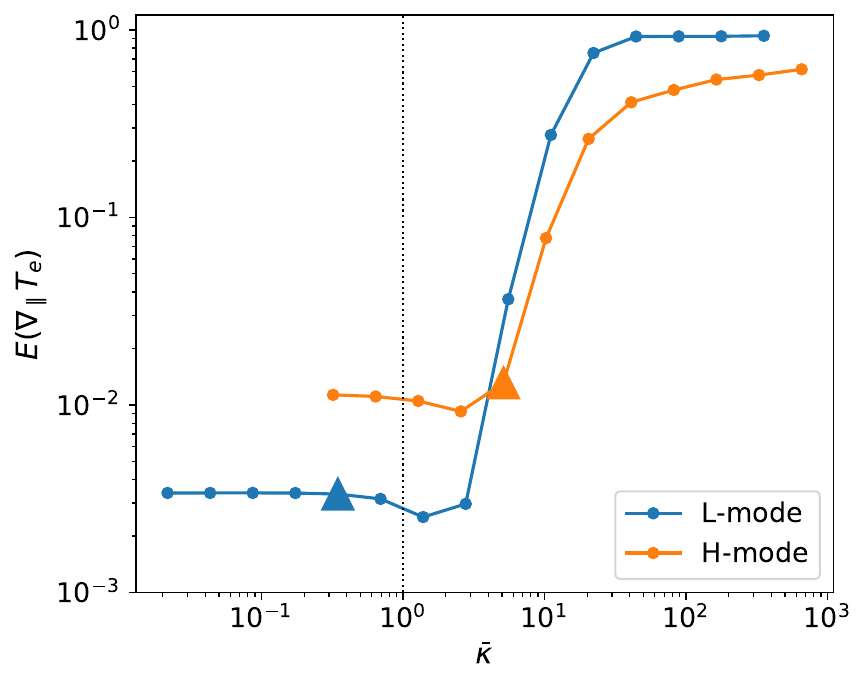}
    \caption{\label{fig:flutter_approx_error_comp}}
\end{subfigure}
\caption{\label{fig:flutter_approx_comp} (a) The parallel gradient of $T_e$ along the perturbed field line scales with $\hat{\kappa}=B_s h_\parallel/ (B_0 h_\perp)$ at one point ($R/R_0=1.2$, $Z/R_0=-0.2$), comparing the approaches of flutter and retracing. $B_s$ is amplified by $n_B$.
(b) The global error \eqref{eqn:error} of flutter terms scales with $\bar{\kappa}=n_B q_{95} \beta_{95} h_\parallel/h_\perp$, comparing the L-mode and H-mode simulations. Triangles denote the actual points in simulations at $n_B=1$.}
\end{figure}

Now we try to demonstrate the above theoretical outcomes in GRILLIX simulations and discuss the possibility of merging the magnetic shift into flutter terms.
In a saturated turbulence run, we consider the temporal and toroidal average $\langle A_\parallel \rangle_{t, \varphi}$ as the magnetic shift $A_s$ which drives that perturbation $ \mathbf{B}_s=\nabla\times A_s \mathbf{b}_0$.
While we freeze the turbulent field of $T_e$, the magnitude of $A_s$ is artificially multiplied by a factor $n_B$, which is scanned from $2^{-4}$ to $2^{10}$ logarithmically (originally $n_B=1$).
Then, the full parallel gradient is computed in two different ways:
(1) Take $n_B \times (2 \pi R A_s) $ as the perturbation to the poloidal magnetic flux $\Psi$ in equilibrium (assuming $A_s$ is purely toroidal), and trace the perturbed magnetic field line based on $\Psi +2\pi n_B  R A_s $.
Then we construct the parallel operator directly from the perturbed field line coordinate, denoted $\nabla_\parallel^{\mathrm{retrace}}$. 
(2) keep the field line coordinate unchanged, and add the flutter term $\nabla^{\mathrm{flutter}}$ (approximated by the right-hand side of \eqref{eqn:flutter_approx}) to the equilibrium parallel gradient $\nabla_\parallel^{\mathrm{equi}}$.
Ideally, $\nabla_\parallel^{\mathrm{retrace}}-\nabla_\parallel^{\mathrm{equi}}$ should be equivalent to the left hand side of \eqref{eqn:flutter_approx} and represent the ideal result of $\nabla^{\mathrm{flutter}}$.
Fig.~\ref{fig:flutter_approx_single} compares $\nabla_\parallel^{\mathrm{retrace}}$ and $\nabla_\parallel^{\mathrm{equi}}+\nabla^{\mathrm{flutter}}$ in an H-mode simulation \cite{zholobenko2024tokamak} at the point ($R/R_0=1.2$, $Z/R_0=-0.2$).
In this run, the perpendicular resolution is $h_\perp=1.9\rho_s$. 
The 16 poloidal planes are used, resulting in a parallel grid length $h_\parallel \approx 2\pi R_0/16$.
The x-axis is scaled by $\hat{\kappa}$, which is determined by the local $B_s$ according to \eqref{eqn:flutter_limit_nom}.
At $\hat{\kappa}=0$ is the parallel gradient in the unperturbed equilibrium $\nabla_\parallel^{\mathrm{equi}} T_e$.
At small $\hat{\kappa}$, there is a linear range where $\nabla_\parallel^{\mathrm{retrace}}$ coincides with $\nabla_\parallel^{\mathrm{equi}}+\nabla^{\mathrm{flutter}}$. 
As $\hat{\kappa}$ increases, $\nabla_\parallel^{\mathrm{retrace}}$ becomes arbitrary, depending on the turbulent structures, while $\nabla^{\mathrm{flutter}}$ increases linearly towards infinity.
As a result, the error between the two diverges.
The current location is strongly affected by turbulence on the low field side. 
Hence, we anticipate $l_\perp$ hitting the resolution $l_\perp \approx h_\perp$ and $\hat{\kappa}\approx \kappa$.
The transition from linear to nonlinear ranges starts at approximately $\hat{\kappa}=1$, verifying the requirement \eqref{eqn:flutter_limit}.
However, at a different location where the turbulence is weaker and $l_\perp > h_\perp$, we find that the linear range of $\hat{\kappa}$ could be much longer than $1$ due to the possibility of $\kappa<1<\hat{\kappa}$.
In that case, one may get an overoptimistic local estimation of the linear validity, which is probably false somewhere else.
To avoid sampling bias, we integrate and normalize the error of flutter results globally
\begin{equation}\label{eqn:error}
   E({\nabla_\parallel T_e})=\frac{\iiint|(\nabla_\parallel^\mathrm{retrace}T_e-\nabla_\parallel^\mathrm{equi}T_e) - \nabla^\mathrm{flutter}T_e |^2 \mathrm{d}V}{\iiint|\nabla^\mathrm{flutter}T_e|^2 \mathrm{d}V}\, ,
\end{equation}
with $\iiint \mathrm{d}V$ being the global volume integral (excluding the buffer zone and the penalization boundary).
Then we need to estimate the order of $B_s/B_0$ assuming that $B_s$ is exclusively caused by the Shafranov shift (PS current associated part).
With a small inverse aspect ratio $a/R_0$ ($\approx0.3$ in AUG), the equation (2.57) in \cite{zohm2014magnetohydrodynamic}  shows that the poloidal magnetic field is perturbed by the Shafranov shift with the scaling ${B_s}/B_\mathrm{pol}\sim \beta_\mathrm{pol} {a}/{R_0}$ (the poloidal beta is defined as $\beta_\mathrm{pol}=2\mu_0p/B_\mathrm{pol}^2$).
Considering $B_0/B_\mathrm{pol}\sim q R_0/a$, we obtain ${B_s}/B_0\sim q \beta$, where $\beta=2\mu_0p/B_0^2$.
As an estimation, we insert the flux surface average of pressure $\langle p \rangle_{\theta}$ to $\beta_{95}=2\mu_0\langle p \rangle_{\theta}/B_0^2$ and take the safety factor $q_{95}$ at $\rho_\mathrm{pol}=0.95$, and rewrite \eqref{eqn:flutter_limit_nom} as   
\begin{equation}\label{eqn:flutter_limit_global}
    \bar{\kappa}=q_{95} \beta_{95} h_\parallel/h_\perp
\end{equation}
In the H-mode simulation \cite{zholobenko2024tokamak}, we measures $q_{95}=3$, $\beta_{95}=0.004$, $h_\parallel\approx2\pi R_0/16$ and $h_\perp\approx R_0/1142$, leading to $\bar{\kappa}=5.12$.
In the L-mode simulation \cite{eder2024self}, they are $q_{95}=4$, $\beta_{95}=0.0018$, $h_\parallel\approx2\pi R_0/16$ and $l_\perp\approx R_0/1142$, leading to $\bar{\kappa}=0.35$.
Assuming that the smallest $l_\perp$ in the whole domain of simulation hits $h_\perp$, the above estimation $\bar{\kappa}$ will be approximately the supremum of $\kappa$, namely $\bar{\kappa}\approx \max(\kappa)$.
Then, based on the theory \eqref{eqn:flutter_limit}, we suspect that the flutter terms are probably able to handle the magnetic shift in L-mode where $\bar{\kappa}=0.35$, but not in H-mode where $\bar{\kappa}=5.12$. 
To verify this guess, we scan the error \eqref{eqn:error} by amplifying $n_B\mathbf{B}_s$ (therefore $\bar{\kappa}=n_B \times$ \eqref{eqn:flutter_limit_global}), and the results are shown in Fig.~\ref{fig:flutter_approx_comp}.
While the magnitudes of error can be affected by numerical discretization, a good proxy of the validity of flutter approximation is the $\bar{\kappa}$-dependence of the error.
The flutter term is valid in the linear range, where both sides of \eqref{eqn:flutter_approx} are proportional to $B_s$.
Therefore $B_s$ should be cancelled in \eqref{eqn:error} and there must be ${\partial E}/{\partial \bar{\kappa}}=0$.
For the L-mode case, the scan shows that the original point $\bar{\kappa}=0.35$ (denoted by the triangle) stays within the flattened range where ${\partial E}/{\partial \bar{\kappa}}$ is zero.
By contrast, the H-mode case with $\kappa=5.12$ is in the transition range with a strong ${\partial E}/{\partial \bar{\kappa}}$, which suggests that the flutter approximation is not valid anymore. 
For both cases, $\bar{\kappa}=1$ serves as a very good prediction for the start of the transition, verifying the success of the theory \eqref{eqn:flutter_limit_global} and \eqref{eqn:flutter_limit}.


\section{Design of a real-time filter for magnetic fluctuations}\label{sec:method}
Integrating a strong magnetic shift into flutter terms is problematic in H-mode conditions, as shown in section~\ref{sec:limit}.
Therefore, it is necessary to keep the magnetic shift in the equilibrium magnetic field, ensuring that magnetic fluctuations remain a small delta-\textit{f} quantity.
In practice, because the background pressure also evolves in full-$f$ simulations, we need to dynamically extract the delta-\textit{f} part of $A_\parallel$, denoted $A_1$, and then compute the magnetic fluctuations via $\mathbf{B}_1=\nabla \times A_1 \mathbf{b}_0$ during the simulation.
To achieve this, we design a real-time filter that only passes the high-frequency component of $A_\parallel$, while tracking the low-frequency
with a new variable $A_s$. The latter is evolved continuously at each time step via
\begin{equation}\label{eqn:evol_apar}
\frac{\partial A_s}{\partial t} = \frac{A_\parallel - A_s}{\lambda},
\end{equation}
where $\lambda$ is a non-negative parameter with a unit of time.
In tokamaks, the background magnetic field is axisymmetrical, so the toroidal average $\langle A_{s} \rangle_\varphi$ is subtracted additionally, yielding
\begin{equation}\label{eqn:evol_A_1}
A_1 =A_\parallel - \langle A_s \rangle_\varphi.
\end{equation}
In stellarators, this toroidal average is impractical and omitted.

With a prescribed arbitrary initial value $A_s (t_0,\lambda)=A_{s,0}$, the analytical solution to \eqref{eqn:evol_apar} reads
\begin{equation}\label{eqn:A_s_abr}
    A_s(t,\lambda)  = e^{-(t-t_0)/\lambda} A_{s,0} 
                    + \left( 1-e^{-(t-t_0)/\lambda} \right) \frac{\int_{t_0}^{t} e^{-(t-\tau)/\lambda} A_\parallel (\tau) \mathrm{d} \tau}{\int_{t_0}^{t} e^{-(t-\tau)/\lambda} \mathrm{d} \tau} \, ,
\end{equation}
where $A_\parallel(\tau)$ is convolved in time.
It is noticed that Eqn.~\eqref{eqn:evol_apar} converges to
\begin{equation}
    \lim_{\lambda\rightarrow +0} A_s(t,\lambda)=A_\parallel(t) \,
\end{equation}
and 
\begin{equation}
    \lim_{\lambda\rightarrow +\infty} A_s(t,\lambda)=A_{s,0}.
\end{equation}

The $C^0$ smoothness of $A_s$ in time is of great importance, as the discontinuity of $A_s$ will trigger artificial instability in the simulations and impede the saturation \cite{zholobenko2024tokamak}.
With a fixed $\lambda$, \eqref{eqn:A_s_abr} $A_s$ has at least $C^1$ smoothness in time.
Even with an abruptly changed $\lambda$, the $C^0$ smoothness of $A_s$ can be guaranteed, as long as the initial values $A_{s,0}$ are identical to the last $A_{s}$ before changing $\lambda$, implied by \eqref{eqn:A_s_abr}.
Hence, it is practical to adjust $\lambda$ during the run.

The Eqn.~\eqref{eqn:evol_apar} and \eqref{eqn:evol_A_1} together serve as a high-pass filter in the time domain. As an output, $A_1$ will possess only the high-frequency ($\gtrsim \lambda^{-1}$) part of $A_\parallel$, while the low-frequency ($\lesssim \lambda^{-1}$) and stationary background enters $A_s$ and is removed.
The main idea of Eqn.~\eqref{eqn:evol_apar} is to keep the background of $A_\parallel$ continuously updated in $A_s$.
This can be understood if we look at the solution of Eqn.~\eqref{eqn:A_s_abr} from the perspective of an exponentially weighted average in time, i.e. weighting a recent $A_\parallel$ more, but an old $A_\parallel$ less, tuned by the parameter $\lambda$.
The exponential weighted average, firstly conceptualized by \cite{roberts2000control}, provides a reliable dynamic mean \cite{lucas1990exponentially}, which can quickly recover from the disturbance and remain smooth \cite{pozzi2012exponential}.
It has been extensively used in forecasting \cite{muth1960optimal,holt2004forecasting} and quality control \cite{hunter1986exponentially,ross2012exponentially}.

In terms of implementation, although Eqn.\eqref{eqn:A_s_abr} provides an exact solution, it is highly memory-intensive due to the requirement to store $A_\parallel$ at each time step.
Therefore, rather than implementing Eqn.\eqref{eqn:A_s_abr} directly, we advance Eqn.~\eqref{eqn:evol_apar} in time synchronizing $A_s$ with the time-stepping of the plasma fields. 
The 3rd order Karniadakis time stepping scheme is employed to solve Eqn.~\eqref{eqn:evol_apar} implicitly: $A_s^{t+\Delta t}$ at the time step $t+\Delta t$ is solved according to
\begin{equation}\label{eqn:dis_apar_shift}
    \left(1 + \frac{6\Delta t}{11}\frac{1}{\lambda} \right) A_s^{t+\Delta t}=\frac{18}{11}A_s^{t}-\frac{9}{11}A_s^{t-\Delta t}+\frac{2}{11}A_s^{t-2\Delta t}+\frac{6\Delta t}{11}\frac{1}{\lambda} {A_\parallel^{t+\Delta t}}
\end{equation}
where $A_\parallel^{t+\Delta t}$ has been obtained by advancing Ohm's law \ref{eqn:ohm} prior to Eqn.~\eqref{eqn:dis_apar_shift}.

\begin{figure}[!htbp]
\centering
\includegraphics[width=0.9\linewidth]{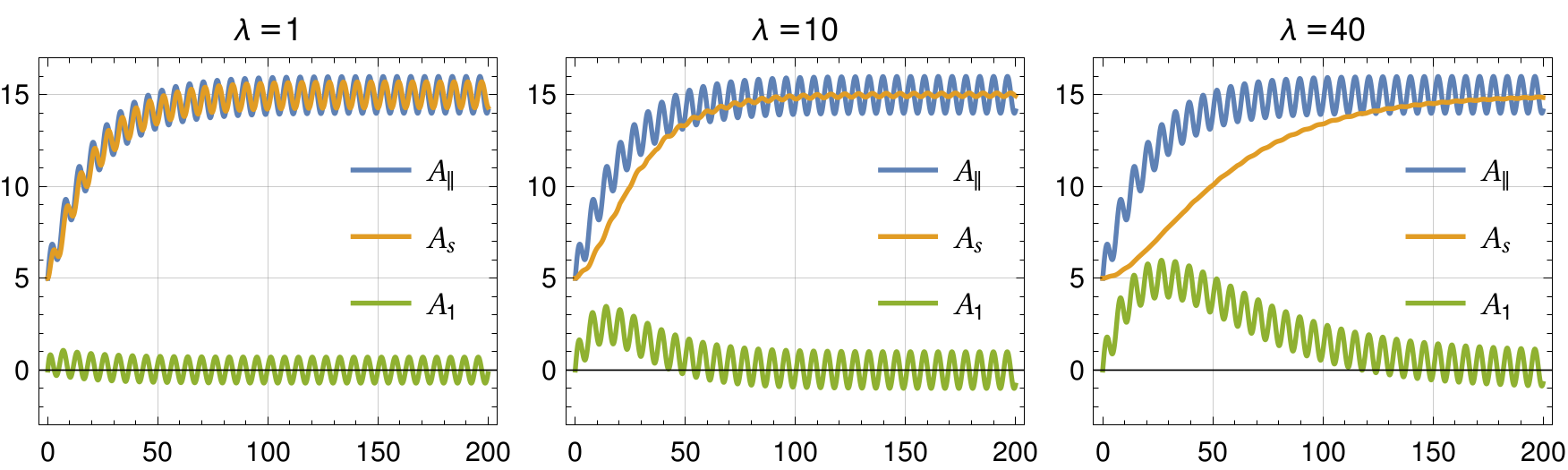}
\caption{\label{fig:unit_test_1} The responses of $A_s$ (Eqn.~\eqref{eqn:unit_ana_As}) to $A_\parallel$ (Eqn.~\eqref{eqn:unit_ana_Apar}) and the resulting $A_1=A_\parallel-A_s$.
$\lambda$ is scanned at $\lambda=1$ (left), $\lambda=10$ (middle), and $\lambda=40$ (right).
The other parameters are fixed as $c_0=5, c_1=10, c_2=1, T=20, \omega=1$.
}
\end{figure}

Before applying the filter to simulations, we try to elaborate the effect of $\lambda$ in this filter analytically.
We manufacture a temporal signal of $A_\parallel$ of the form
\begin{equation}\label{eqn:unit_ana_Apar}
A_\parallel(t) =  c_0 + c_1 \left( 1-e^{- t/T} \right) + c_2 \sin \left( \omega {t} \right), \qquad t \ge 0.
\end{equation}
The first term $c_0$ introduces an initial shift in $A_\parallel$. 
The second term led by $c_1$ describes a gradual increase towards saturation, representing the slow evolution of the background $A_\parallel$ consistent with the evolving pressure. 
The third term led by $c_2$ mimics the turbulent oscillations with high frequency.
Given the initial condition $A_s(0,\lambda)=A_\parallel(0)$, $A_s$ can be obtained analytically 
\begin{equation}\label{eqn:unit_ana_As}
\begin{aligned}
A_s(t,\lambda) & =  c_0 + c_1 \left(1-\frac{T e^{-t/T}- \lambda e^{-t/\lambda}}{T-\lambda} \right)
+ c_2\left(\frac{ \lambda \omega  e^{-t/\lambda}-\lambda \omega  \cos \left(\omega t\right)+\sin \left(\omega t\right)}{\lambda ^2\omega^2+1}\right)\,.
\end{aligned}
\end{equation}
The effect of $\lambda$ is illustrated in Fig.~\ref{fig:unit_test_1}.
When $\lambda=1$, $A_s$ closely follows $A_\parallel$ with minor delay, but retaining the majority of the high-frequency fluctuations. As a result, the fluctuation amplitudes $A_1$ are artificially reduced. 
As $\lambda$ increases, $A_s$ becomes smoother, and the high-frequency components of $A_\parallel$ are better preserved in $A_1$, with a price of higher overshoot at early stages.
In actual turbulence simulations, we are concerned with both the initial overshoot and the preservation of high-frequency components.
The initial overshoot should be small enough that the magnetic perturbation caused by $A_1$ is below the numerical threshold \eqref{eqn:flutter_limit}.
The high-frequency components of $A_\parallel$ must remain as intact as possible in $A_1$ to avoid distorting the turbulent dynamics.
Analytically, the high-frequency component retained in $A_1$ in the limit of $1/(\omega\lambda)\rightarrow0$ is
\begin{equation}\label{eqn:lambda_large}
    \underbrace{ \sin(\omega t)}_{\text{in}\, A_\parallel} \rightarrow \underbrace{\frac{1}{1 + (\omega\lambda)^{-2}}\sin \left(\omega t+\arctan \frac{1}{\omega\lambda} \right)}_{\text{in}\, A_1}
\end{equation}
The distortion can be eliminated if $\omega\lambda \gg 1$, meaning that the minimal $\lambda$ is constrained by the inverse mode frequency.
In this paper, we set the constrain as 
\begin{equation}\label{eqn:lambda_limit}
    \lambda > 10/\omega\, ,
\end{equation}
ensuring the preserving rate of the amplitude $1/(1 + (\omega\lambda)^{-2})>99\%$. We will demonstrate that \eqref{eqn:lambda_limit} serves as a useful estimate for the minimal required $\lambda$, below which the turbulence gets distorted by the filter in simulations. 

\section{Verification in turbulence simulations for tokamaks}\label{sec:sim}
\subsection{L-mode}
\begin{figure}[!htbp]
\centering
\includegraphics[width=0.4\linewidth]{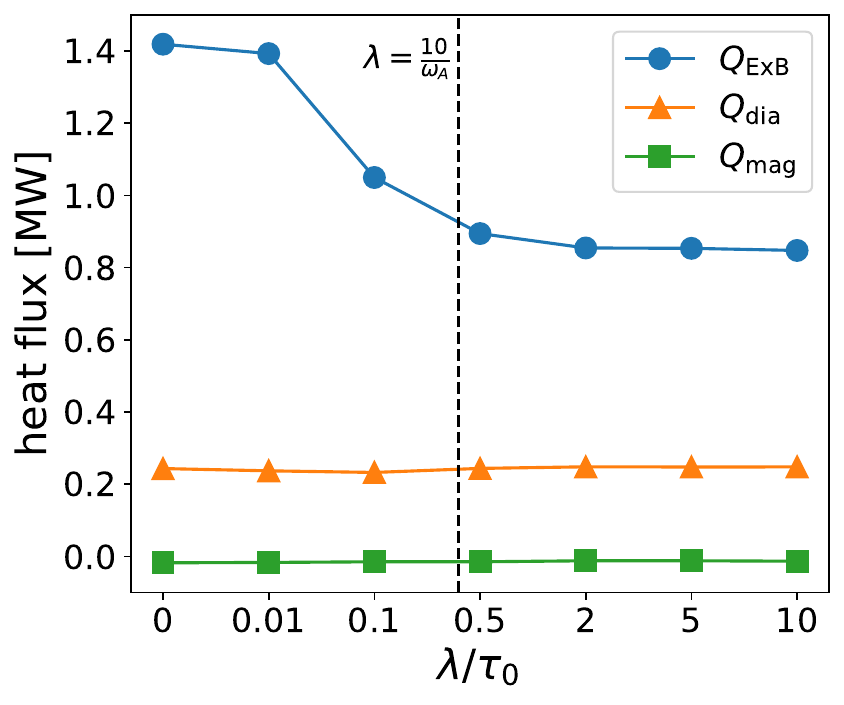}
\caption{\label{fig:lmode_exb_scan} 
The heat fluxes across the flux surface $\rho_\mathrm{pol}=0.975$ as a function of $\lambda$ (x-axis is not linear), averaged in time over the same period.
$\lambda=0$ means removing the instantaneous toroidal average $\langle A_\parallel\rangle_\varphi$.
The components $Q_\mathrm{ExB}$, $Q_\mathrm{dia}$ and $Q_\mathrm{mag}$ (expressions in Eqn.(13) and Eqn.(14) in \cite{zhang2024magnetic}) are $E\times B$ heat flux, diamagnetic heat flux, and magnetic flutter heat flux, respectively, counting in both ions and electrons.
The vertical dash line depicts the `$10/\omega$' rule \eqref{eqn:lambda_limit} with $\omega_A\approx 10^3$kHz.}
\end{figure}

The background $A_\parallel$ will not violate the flutter approximation in L-mode conditions, as discussed in section \ref{sec:limit}. 
The main motivation for filtering out the background $A_\parallel$ is to ensure no overlap between $\nabla_{\parallel}^\mathrm{equi}$ and $\nabla^\mathrm{flutter}$, improving physical consistency.
While the low-frequency component of $A_\parallel$ does not risk exceeding the numerical limit \eqref{eqn:flutter_limit}, we need to use the filter with caution to avoid the probable distortion of the high-frequency component.
Starting from a saturated turbulence stage at $t=224\tau_0$ ($\tau_0=23.8 \ \mu\mathrm{s}$), we scan the filter parameter $\lambda$ at $0$, $0.01\tau_0$, $0.1\tau_0$, $0.5\tau_0$, $2\tau_0$, $5\tau_0$, and $10\tau_0$, and initialize $A_s = A_\parallel$ for all cases. 
Note that using $\lambda=0$ is equivalent to removing the instantaneous toroidal average $\langle A_\parallel \rangle_\varphi$ at each time step.
The simulations employ the adaptive heat source to maintain the temperatures at the core boundary $\rho_\mathrm{pol}=0.9 \sim 0.92$ \cite{zholobenko2021electric}.
The changes in turbulent dynamics will cause a different heating power injected into the system, leading to different transports across the edge.
In L-mode, the edge transport is primarily driven by $E \times B$ dynamics \cite{zhang2024magnetic}. 
Our simulations show that $E \times B$ transport is highly sensitive to $\lambda$, while diamagnetic transport is unaffected and flutter transport remains negligible, which are common for any a flux surface $0.94<\rho_\mathrm{pol}<1$ and illustrated at $\rho_\mathrm{pol}=0.975$ in Fig.~\ref{fig:lmode_exb_scan}.
The cases $\lambda=0,0.01\tau_0$ exhibit notably higher transport levels than the others.
As $\lambda$ increases beyond $0.5\tau_0$, the $E \times B$ transport converges.

\begin{figure}[!htbp]
\centering
\begin{subfigure}[b]{0.4\textwidth}
    \includegraphics[width=\textwidth]{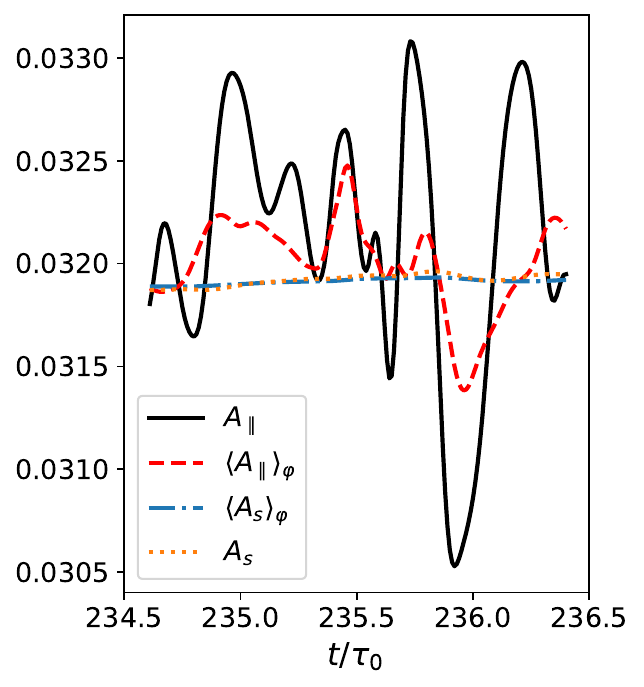}
    \caption{\label{fig:alfv_trace}}
\end{subfigure}
\begin{subfigure}[b]{0.4\textwidth}
    \includegraphics[width=\textwidth]{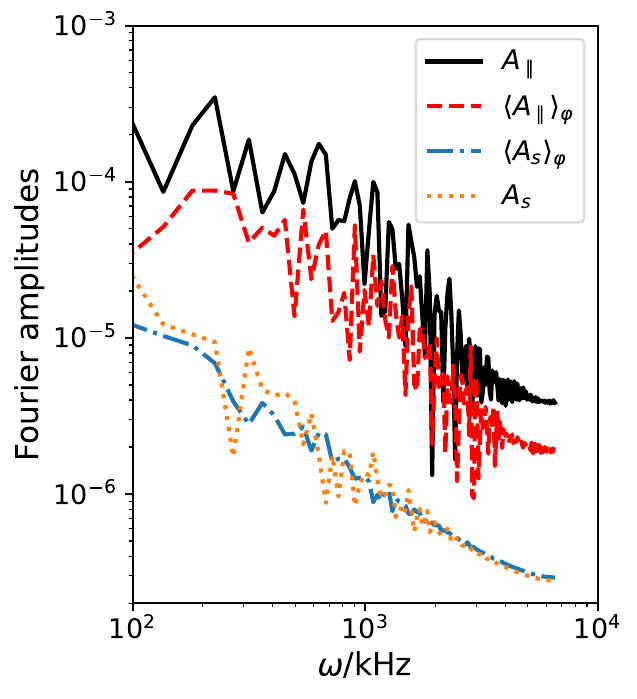}
    \caption{\label{fig:alfv_freq}}
\end{subfigure}
\caption{\label{fig:alfv} (a) The temporal signals of $A_\parallel$, its toroidal average $\langle A_\parallel \rangle_\varphi$, $A_s$, and its toroidal average $\langle A_s \rangle_\varphi$ at one point ($\rho_\mathrm{pol}=0.975$ at outboard-mid plane), with $\lambda=5\tau_0$ and $\tau_0=0.0238 \ \mathrm{ms}$.
(b) The temporal Fourier amplitudes of the above four signals in frequency space. $\omega$ is the circular frequency.
All signals are normalised to $\rho_{s0} B_\mathrm{axis}=1.44\times 10^{-3}\mathrm{m}\cdot\mathrm{T}$.
}
\end{figure}

The analysis of the filter in section \ref{sec:method} suggests that the higher transport observed at $\lambda=0$ (the toroidal average) is a numerical artifact. 
To corroborate this further, fig.~\ref{fig:alfv_trace} compares at a single point ($\rho_\mathrm{pol}=0.975$, outboard mid-plane) in the case $\lambda=5\tau_0$ four temporal signals: the original $A_\parallel$, its toroidal average $\langle A_\parallel \rangle_\varphi$, the magnetic shift $A_s$ evolved according to \eqref{eqn:evol_apar} and its toroidal average $\langle A_s \rangle_\varphi$. 
A clear resonance is visible between $\langle A_\parallel \rangle_\varphi$ and $A_\parallel$.
It is known that magnetic flutter can stabilize $E \times B$ transport in L-mode edge conditions \cite{zhang2024magnetic}. 
Removing $A_1 = A_\parallel - \langle A_\parallel \rangle_\varphi$ dampens the amplitudes of $A_1$ and subsequent flutter terms by canceling the resonant components between $A_\parallel$ and $\langle A_\parallel \rangle_\varphi$, which reduces the stabilizing effect of flutter and leads to increased $E \times B$ transport.

To determine the resonance frequency range, Fig.~\ref{fig:alfv_freq} shows the Fourier spectra of the above four signals. Resonance begins around $\omega = 10^3 \ \mathrm{kHz}$, where the red ($\langle A_\parallel \rangle_\varphi$) and black ($A_\parallel$) lines converge. This aligns with the Alfvén frequency, estimated as $\omega_A \approx k_\parallel v_A = 10^3 \ \mathrm{kHz}$, given the local Alfvén speed $v_A \approx B/\sqrt{\mu n_i m_i} = 10^7 \ \mathrm{m/s}$ and $k_\parallel = 1/(q R_0)$.
In contrast, $\langle A_s \rangle_\varphi$ eliminates Alfv\'en oscillations with Fourier amplitudes an order of magnitude below those of the original $A_\parallel$. The amplitude of $A_s$ without toroidal averaging is similar to that of $\langle A_s \rangle_\varphi$, both well-separated from $A_\parallel$, indicating that toroidal averaging is less effective at filtering out Alfv\'en oscillations than time integration in $A_s$. 
This is because Alfv\'en waves propagate along magnetic field lines, which are not aligned with the toroidal direction due to helicity. A sufficient time average can effectively filter out Alfv\'en oscillations independently of field line topology.
According to equation \eqref{eqn:lambda_limit}, the filter can achieve this with $\lambda > 10/\omega_A \approx 0.4 \tau_0$. 
This criterion explains why in Fig.~\ref{fig:lmode_exb_scan}, $\lambda=0.01 \tau_0$ results in similar artificially high transport as the $\lambda=0$ case, while the case with $\lambda=0.5 \tau_0$ yields a transport level close to the converged value.

\subsection{H-mode}
\begin{figure}[!htbp]
\centering
\begin{subfigure}[b]{0.4\textwidth}
    \includegraphics[width=\textwidth]{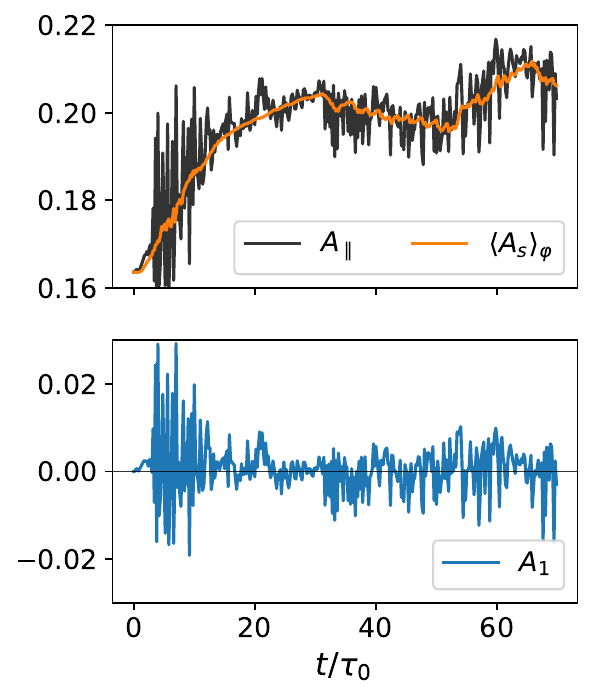}
    \caption{\label{fig:trace_apar_1}}
\end{subfigure}
\begin{subfigure}[b]{0.4\textwidth}
    \includegraphics[width=\textwidth]{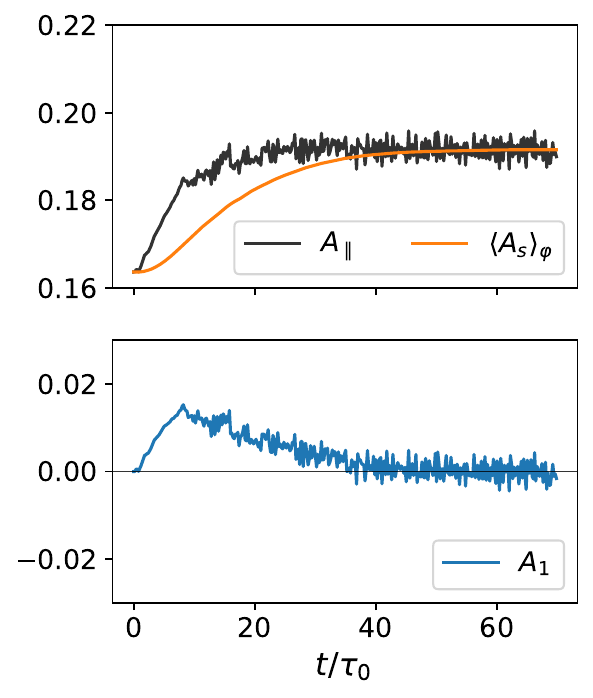}
    \caption{\label{fig:trace_apar_10}}
\end{subfigure}
\caption{\label{fig:tracing}The dynamic response of the $\langle A_s\rangle_\varphi$ to $A_\parallel$ and the resulting $A_1=A_\parallel-\langle A_s\rangle_\varphi$ with (a)$\lambda=1\tau_0$ and (b)$\lambda=10\tau_0$. All signals are measured at one point ($\rho_\mathrm{pol}=0.996$ at outboard-mid plane), normalised to $\rho_{s0} B_\mathrm{axis}=1.44\times 10^{-3}\mathrm{m}\cdot\mathrm{T}$.}
\end{figure}

We conclude from the L-mode tests that the parameter $\lambda$ must be sufficiently large to prevent interference with the Alfvén waves. Under H-mode conditions, the filter is subjected to more stringent challenges.
On one hand, the slowly evolving background part of $A_\parallel$ can numerically violate the flutter approximation, as illustrated in Fig.~\ref{fig:flutter_approx_error_comp}, requiring a finite $\lambda$. 
On the other hand, we will show below that the turbulent dynamics require $\lambda > 3\tau_0$. 
Our investigations are based on previous H-mode simulations of AUG discharge \#40411 \cite{zholobenko2024tokamak}, where originally an inferior method for dealing with the magnetic shift has been used. Starting from $t=0$, we evaluate the performance of our new filter with varying $\lambda$.
Fig.~\ref{fig:tracing} shows the responses of $\langle A_s \rangle_\varphi$ to $A_\parallel$ with $\lambda=1\tau_0$ and $\lambda=10\tau_0$.
As anticipated, with $\lambda=1\tau_0$, $\langle A_s \rangle_\varphi$ more closely follows $A_\parallel$, remaining consistently centered within the fluctuations of $A_\parallel$.
However, in this filter setting, $\langle A_s \rangle_\varphi$ exhibits rapid fluctuations that resonate with $A_\parallel$, artificially amplifying the amplitude of $A_1$.
In contrast, the $\lambda=10\tau_0$ case shows a noticeable deviation between $\langle A_s \rangle_\varphi$ and the center of $A_\parallel$ during the initial phase.
As $A_\parallel$ gradually saturates, $\langle A_s \rangle_\varphi$ increases smoothly, eventually stabilizing as a constant background for $A_\parallel$.
Despite the initial deviation, the maximal amplitudes of $A_1$ are approximately an order of magnitude lower than the original $A_\parallel$.
As demonstrated in Fig.~\ref{fig:flutter_approx_error_comp}, decreasing $\bar{\kappa}$ by a factor of 10 shifts the original triangle to the `flat' region where the flutter approximation is valid. 
Thus, applying the filter with $\lambda=10\tau_0$ enables the H-mode simulation to proceed without encountering numerical instabilities.

\begin{figure}[!htbp]
\centering
\begin{subfigure}[b]{0.4\textwidth}
\includegraphics[width=\textwidth]{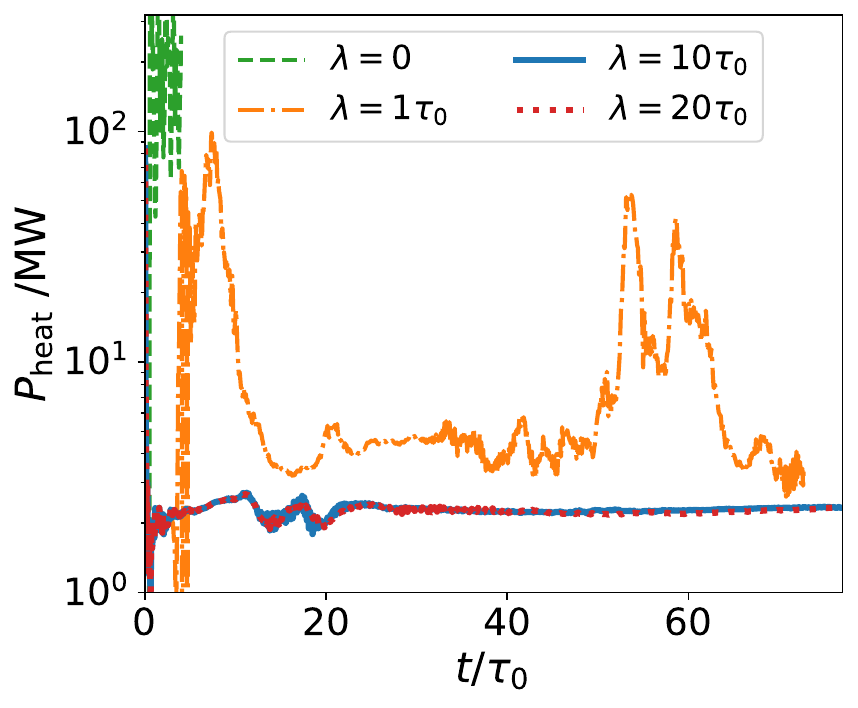}
\caption{\label{fig:tracing_P_hmode}}
\end{subfigure}
\begin{subfigure}[b]{0.4\textwidth}
\includegraphics[width=\textwidth]{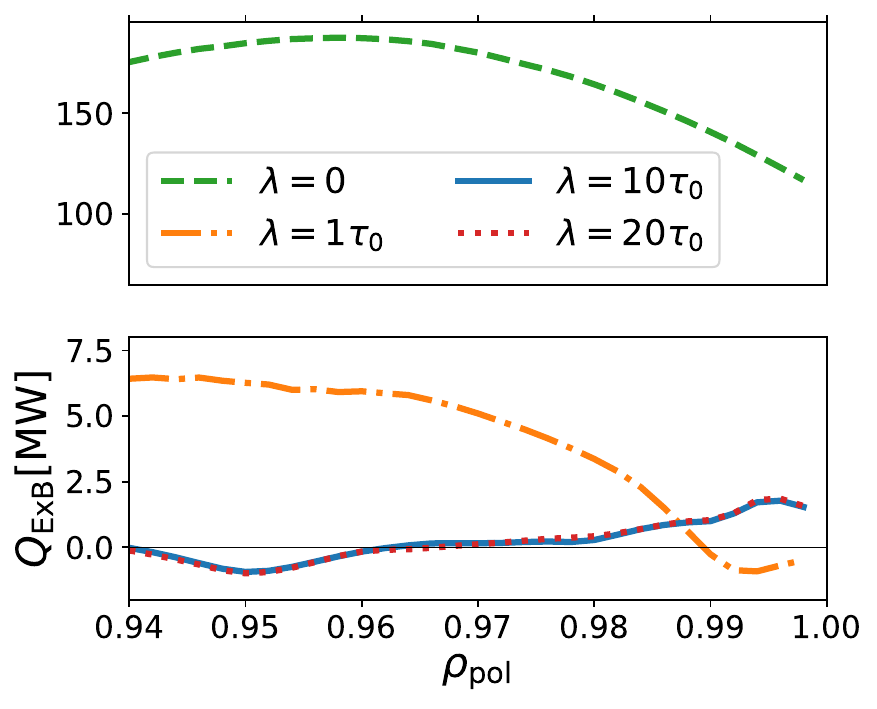}
\caption{\label{fig:hmode_exb_rho}}
\end{subfigure}
\caption{\label{fig:hmode_scan}(a) The time evolution of the heating power, scanning $\lambda=0, 1\tau_0, 10\tau_0, 20\tau_0$ (b) The surface integral of $E\times B$ heat flux, averaged in time over $t=2\sim4\tau_0$ for the case $\lambda=0$ and over $t=40\sim70\tau_0$ for the cases $\lambda=1\tau_0, 10\tau_0, 20\tau_0$.}
\end{figure}

To examine the effect of $\lambda$ on the transport, we scan the parameter $\lambda=0, 1, 10, 20\tau_0$ and track the total heating power $P_\mathrm{heat}$ injected by the adaptive heat source.
The value of $P_\mathrm{heat}$ indicates the turbulent transport level in our simulations.
Fig.~\ref{fig:tracing_P_hmode} illustrates the time evolution of $P_\mathrm{heat}$ starting from $t=0$.
As $\lambda$ increases, the heating power decreases and converges to approximately $2.2$MW.
The results for $\lambda=10\tau_0$ and $\lambda=20\tau_0$ are almost identical.
After turbulence saturates at approximately $60\tau_0$, the mean profile remains unchanged.
At this stage, we turn $\lambda$ from $10\tau_0$ to infinity (equivalent to freezing $\langle A_s \rangle_\varphi$), without altering the turbulent transport.
This demonstrates that a filter setting of $\lambda=10\tau_0$ provides a converged outcome.
Therefore we select the case $\lambda=10\tau_0$ as the reference for the following discussions.
Compared to our previous simulations \cite{zholobenko2024tokamak}, such a filter with an appropriately chosen width ($\lambda=10, 20\tau_0$) indeed accelerates the saturation.

It is useful to analyze the issues associated with using $\lambda=0$ (instantaneous toroidal average) and $\lambda=1\tau_0$ (insufficient filter width) as further references to better understand the appropriate filter width setting.
For $\lambda=0$, the system experiences an extremely high heating power, approximately two orders of magnitude greater than the reference, before crashing in the early stages of the simulation. 
The heating power with $\lambda=1\tau_0$, while lower than that for $\lambda=0$, exhibits strong fluctuations accompanied by burst events, remaining consistently above the reference value.
We find that the discrepancies in heating power among the three cases ($\lambda=0, 1, 10\tau_0$) are primarily driven by $E\times B$ heat transport. 
The diamagnetic transport appears insensitive to the filter setting, while the flutter transport behaves differently but contributes less than 10\% of the total heating power across all three cases. 
Thus, we focus on $E\times B$ heat transport, for which the surface integral $Q_\mathrm{ExB}$ is shown in Fig.~\ref{fig:hmode_exb_rho}. 
For $\lambda=0$, the $E\times B$ transport is consistently higher by more than $100$MW across the domain.
This is reminiscent of the scenario in which the flutter terms are omitted completely, as studied in \cite{zholobenko2024tokamak}, leading to $E\times B$ fluxes that are increased by a factor of $10^2$ due to the absence of flutter stabilization in drift-Alfv\'en waves. 
In this context, the issue with using $\lambda=0$ under H-mode conditions can be understood as analogous to the L-mode scenario: filtering out the Alfv\'en dynamics weakens flutter stabilization, resulting in enhanced $E\times B$ transport.

The issue with using $\lambda=1\tau_0$ is more complex due to its $\rho_\mathrm{pol}$-dependence. 
Between $\rho_\mathrm{pol}=0.94$ and $0.98$, $Q_\mathrm{ExB}$ is higher than the reference value, due to the same mechanism involving the interference of Alfv\'en dynamics. 
Because of a higher plasma density, this H-mode case has a faster Alfv\'en speed than that in L-mode, leading to the typical Alfv\'en frequency of $\omega_A \approx 3\times10^2 \mathrm{kHz}$ in H-mode, compared to $10^3 \mathrm{kHz}$ in L-mode.
This difference likely explains why $\lambda=0.5\tau_0$ works adequately in L-mode (as shown in Fig.~\ref{fig:lmode_exb_scan}) but $\lambda=1\tau_0$ proves problematic under H-mode conditions.
This restriction relaxes as $\rho_\mathrm{pol}$ increases and the density decreases rapidly across the pedestal, leading to an increase in $\omega_A$. 
However, for $\rho_\mathrm{pol}>0.99$ at the pedestal foot, $Q_\mathrm{ExB}$ with $\lambda=1\tau_0$ falls below the reference case, which cannot be attributed to flutter stabilization. 
Instead, this behavior is caused by interactions between the filter and the low-frequency component of the kinetic ballooning mode (KBM) \cite{cheng1981kinetic,tang1980kinetic} at the pedestal foot.

\begin{figure}[!htbp]
\centering
\begin{subfigure}[b]{0.325\textwidth}
    \includegraphics[width=\textwidth]{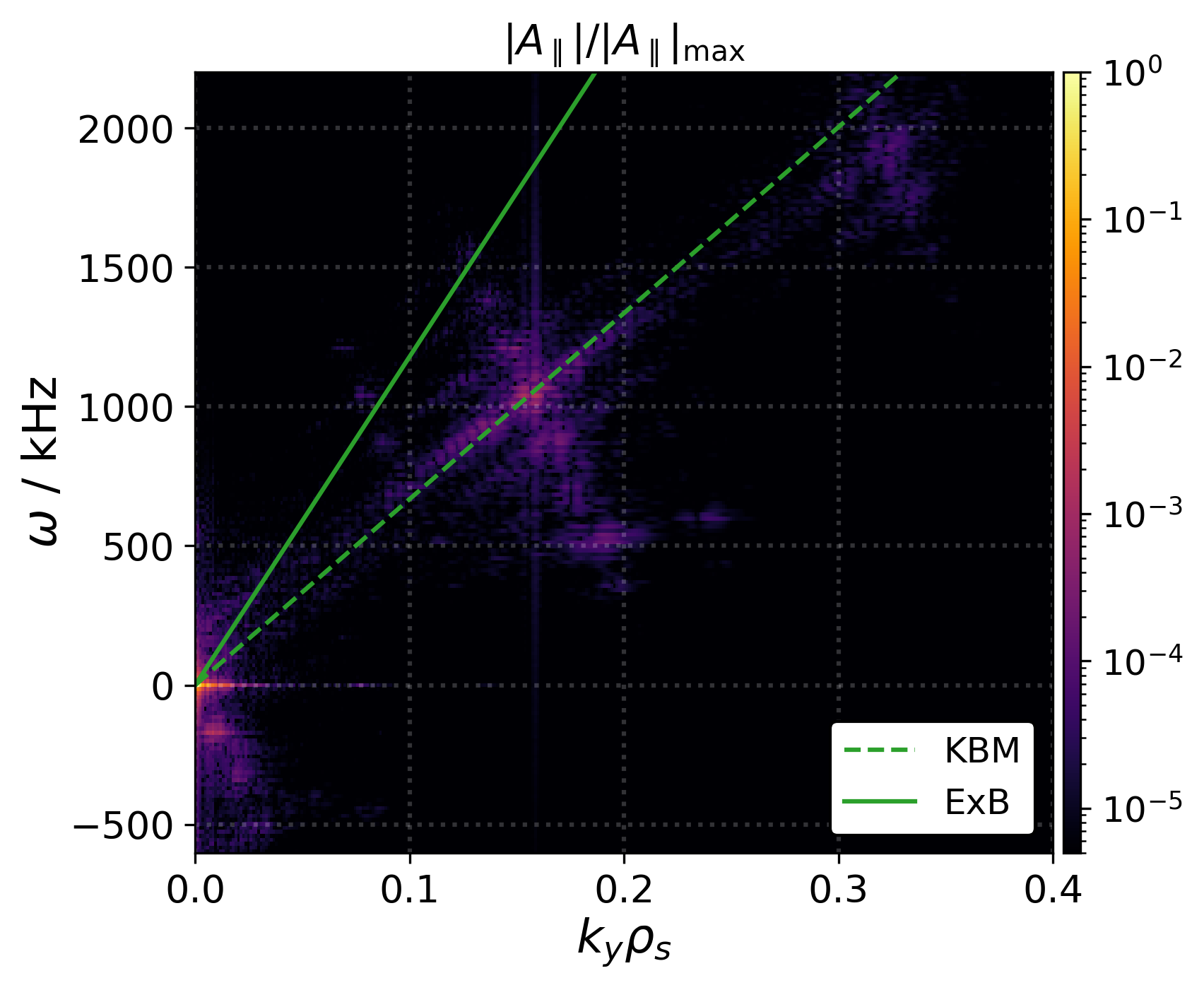}
    \caption{\label{fig:disper_apar_t_10}}
\end{subfigure}
\begin{subfigure}[b]{0.325\textwidth}
    \includegraphics[width=\textwidth]{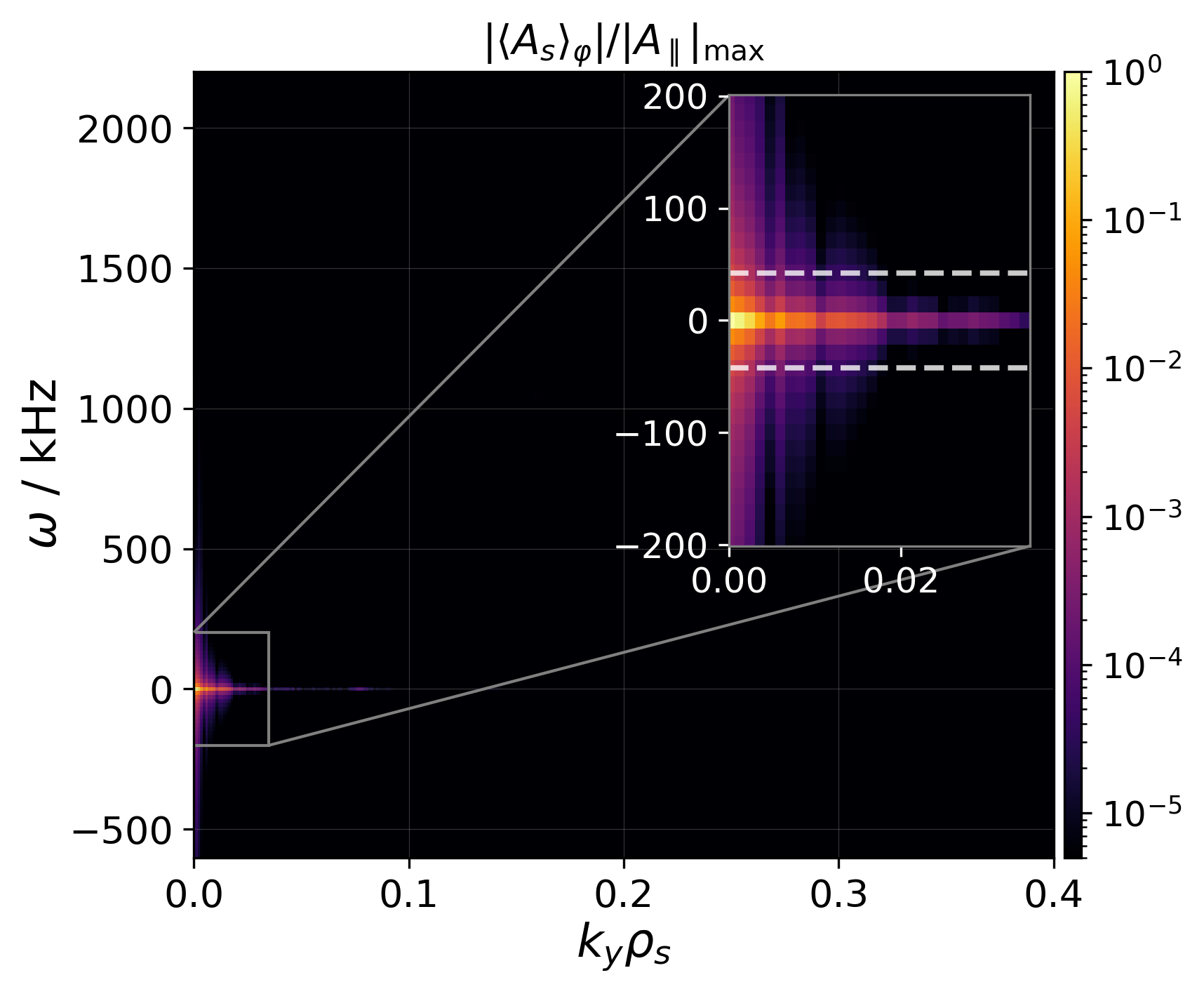}
    \caption{\label{fig:disper_apar_s_10}}
\end{subfigure}
\begin{subfigure}[b]{0.325\textwidth}
    \includegraphics[width=\textwidth]{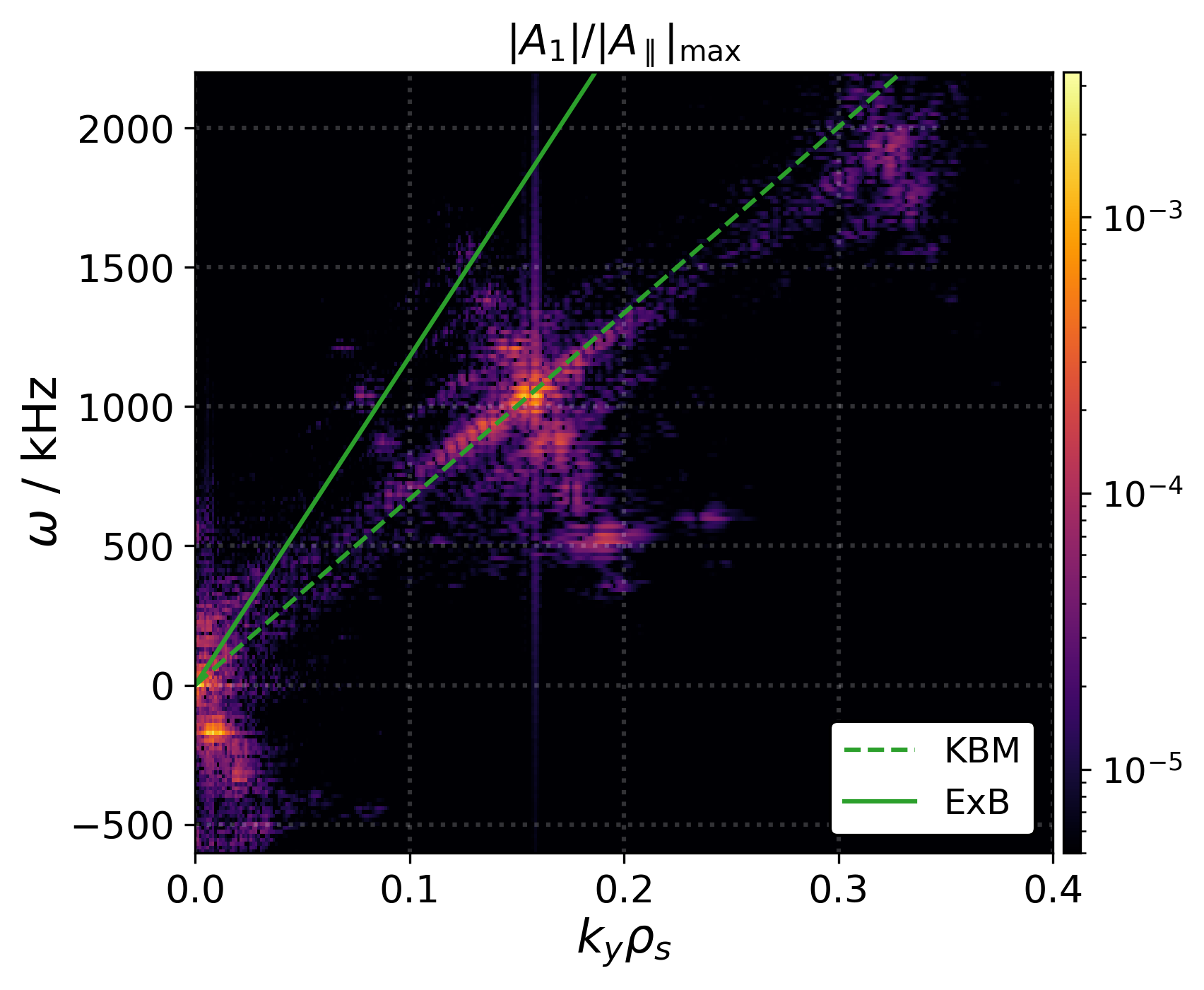}
    \caption{\label{fig:disper_apar_f_10}}
\end{subfigure}
\caption{\label{fig:disper_apar_10}The Fourier spectra of (a) $A_\parallel$, (b) $\langle A_s\rangle_\varphi$, and (c) $A_1$ in the laboratory frame at $\rho_\mathrm{pol}=0.995$ in the $\lambda=10\tau_0$ case. The solid line represents the $E\times B$ rotation direction. The green dashed lines correspond to the KBM desperation on the $E\times B$ rotation basis. }
\end{figure}

The unstable kinetic ballooning mode (KBM) at the pedestal foot imposes a constraint on the minimal filter width in H-mode conditions. 
The relevance of the KBM to the H-mode pedestal has been established in prior studies \cite{Dickinson2012, snyder2011first, zholobenko2024tokamak}. 
The KBM is an electromagnetic ballooning mode driven by the ion temperature gradient through the ion magnetic drift resonance, which can be described using the two-fluid model with the polarization approximation \cite{hirose1995ion}.
Fig.~\ref{fig:disper_apar_t_10} illustrates the Fourier spectra of $A_\parallel$ for the $\lambda = 10\tau_0$ case, plotted with respect to the angular frequency $\omega$ (positive is the electron diamagnetic direction) and $k_y\rho_s$, where $k_y$ represents the poloidal wave number in the symmetry angle \cite{ulbl2023influence} at the flux surface $\rho_\mathrm{pol} = 0.995$. 
The magnetic shift is observed at $\omega = 0$ and $k_y\rho_s \approx 0$, where the Fourier amplitudes of $A_\parallel$ become the strongest ($|A_\parallel| / |A_\parallel|_\mathrm{max} = 1$) and dominate over other spectral components.
A mode with three harmonic components is identified below the $E \times B$ rotation, represented by the solid green line, indicating propagation in the ion diamagnetic direction. 
This mode is identified as KBM, as its harmonic peaks align closely with the linear KBM dispersion $\omega_{p_i}/2$ \cite{aleynikova2017quantitative}, transformed as $\omega_{p_i}/2 + \omega_\mathrm{ExB}$ in the laboratory frame.
In our nonlinear simulations, this KBM exhibits a broad frequency range. At $k_y\rho_s \sim 0.2$, the frequency spans from 500 kHz to 1250 kHz. At longer wavelengths ($k_y\rho_s \sim 0.01$), the frequency even decreases to $-150$ kHz, where the corresponding spectral intensity still remains high. 
To avoid interference with the low-frequency branch at $-150$ kHz, the filter width restriction given in \eqref{eqn:lambda_limit} is $\lambda > 10 / (150\mathrm{kHz}) \approx 3\tau_0$, suggesting that the $\lambda = 1\tau_0$ case may be problematic.

The filter with $\lambda = 10\tau_0$ effectively removes the magnetic shift without significantly affecting the majority of the KBM spectrum. Fig.~\ref{fig:disper_apar_s_10} shows the spectrum of $\langle A_s\rangle_\varphi$ removed by the filter. This spectrum appears as a localized point in Fourier space at $\omega \approx 0$ and $k_y \rho_s \approx 0$, but with high intensity ($|\langle A_s\rangle_\varphi| / |A_\parallel|_\mathrm{max} = 1$), indicating the successful capture of the magnetic shift.
The zoomed-in subplot highlights the rapid decay of spectral intensity along both the $\omega$ and $k_y \rho_s$ directions. The two horizontal dashed lines in the subplot correspond to the `$ \lambda = 10 / \omega$' law described in ~\eqref{eqn:lambda_limit}.
Fig.~\ref{fig:disper_apar_f_10} displays the spectra of $A_1$, which represents the output of the filter, defined as $A_1 = A_\parallel - \langle A_s\rangle_\phi$. The color bar cutoff at $5 \times 10^{-5}$ is consistent with Figure~\ref{fig:disper_apar_t_10}, although the maximum intensity of the spectra decreases by approximately three orders of magnitude. This reduction results in a much clearer depiction of the KBM spectra.
A comparison of Fig.~\ref{fig:disper_apar_f_10} and ~\ref{fig:disper_apar_t_10} confirms that the KBM structure including the low-frequency branches is well preserved by the filter.



\section{Conclusion}
For edge and SOL simulations of magnetic confinement fusion devices, full-\textit{f} turbulence codes evolve the total plasma profile including fluctuations and background. At the same time, because it is computationally beneficial to use field-aligned coordinates, these codes separate the magnetic field into the magnetic equilibrium and perturbations (represented by the parallel magnetic potential $A_\parallel$) which are handled by the magnetic flutter operator.
This study illustrates that a magnetic shift $A_s$, a low-frequency ($\omega_s\ll\omega_t$) component of $A_\parallel$, exists at least in standard full-\textit{f} models. 
This shift may cause physical inconsistency by double counting the Shafranov shift within the magnetic equilibrium and the magnetic perturbations.

The magnetic shift also presents a numerical challenge to field-aligned codes. Unlike turbulent magnetic fluctuations, which manifest as perpendicular drift effects localized within a single poloidal plane, the magnetic shift reconstructs the field-aligned structure macroscopically. The application of the flutter operator to the magnetic shift is problematic, as the magnitude of $B_s$ must satisfy the constraint $\kappa\ll1$ \eqref{eqn:flutter_limit}. 
This limitation has been verified by scanning the magnitude of $B_s$ in L-mode and H-mode turbulence conditions.
Our results show that when $\kappa\ll1$, the magnetic flutter term of $B_s$ agrees with retracing the $B_s$-perturbed field line.
However, for $\kappa \sim 1$, the flutter approximation fails, resulting in deviations from retracing.
Our tests show that the criterion \eqref{eqn:flutter_limit} can be satisfied in the AUG L-mode scenario but not in H-mode.
Therefore, it is desirable to subtract the magnetic shift from flutter and incorporate it into the magnetic equilibrium instead in H-mode simulations.

To filter out the magnetic shift, we propose a high-pass filter \eqref{eqn:evol_apar} and \eqref{eqn:evol_A_1}, tuned by the filter width $\lambda$.
The filter evolves $A_s$ continuously in real-time (at each time step), becoming part of the physical model.
This is in contrast to our previous study \cite{zholobenko2024tokamak} which found that H-mode simulations were very sensitive to discontinuous updates of $A_s$, which triggered artificial instability preventing the turbulence from saturation.
The proposed filter mitigates this issue, accelerating saturation and reducing run times.

The filter width $\lambda$ must be properly chosen to avoid interference with the electromagnetic microinstabilities of interest.
This imposes a lower bound of  $\lambda>10/\omega$, which has been explored in L-mode and H-mode turbulence simulations.
In L-mode, the shear Alfv\'en wave frequency $\omega_A\sim10^3$kHz dictates the restriction. 
Insufficient filter widths $\lambda \lesssim \omega_A$ will weaken the flutter stabilization in drift-Alfv\'en-wave, leading to artificially high $E\times B$ transport.
H-mode simulations are more complex due to the interplay of multiple electromagnetic effects.
While the Alfv\'en wave frequency remains a key constraint, the unstable KBM at the pedestal foot introduces additional restrictions.
The low-frequency branch of KBM around $\omega_\mathrm{KBM}=150\sim500$kHz sets the effective lower bound for $\lambda$.
If $\lambda \lesssim 1/\omega_\mathrm{KBM}$, KBM dispersion is disrupted, compromising the accurate capture of $E\times B$ transport triggered by KBM near the separatrix. 

Importantly, we stress that removing the instantaneous toroidal average $\langle A_\parallel \rangle_\varphi$ is not suitable for field-aligned codes. 
The pure toroidal average fails to smoothen field-aligned modes. This approach pollutes the Alfv\'en wave dynamics, exaggerating turbulence intensity and leading to unrealistic estimates of $E\times B$ transport.
Instead, the present filter emphasizes the time average of $A_\parallel$ indicated by \eqref{eqn:A_s_abr}, while the toroidal average is proved to be unnecessary by Fig.~\ref{fig:alfv}.
This also extends the capability of the filter to the non-axisymmetrical geometry, like stellarators.

While setting $\lambda=\infty$ will minimize the intervention of the filter on the turbulence, this may delay the response of $A_s$ to the evolution of the background $A_\parallel$, increasing the risk of numerical issues due to higher $\kappa$, particularly in H-mode.
In practice, fixing $\lambda=10\tau_0$ allows the H-mode simulation to run through within the numerical constraint \eqref{eqn:flutter_approx}.
At the same time, the turbulent transport and micro instability are not polluted by the filter.
This conclusion is supported by the $\lambda$-scan in Fig.~\ref{fig:tracing_P_hmode} and clean spectra of $A_s$ in Fig.~\ref{fig:disper_apar_s_10}.
Notably, $\lambda$ is not required to be constant throughout a run (the zero order continuity of $A_s$ is ensured even with abrupt changes in $\lambda$, implied by \eqref{eqn:A_s_abr}).
This flexibility enables strategic adjustment of $\lambda$:
smaller values are preferable during the initial phase of rapid background profile evolution, while larger values or even $\lambda=\infty$ can be used after turbulence saturation to ensure fidelity in microinstability dynamics.
In summary, the proposed high-pass filter makes exploring reactor-relevant regimes in high beta scenarios possible for full-\textit{f} field-aligned codes. 

\section{Outlook}
While in our present work we were dealing with removing remnants of the equilibrium magnetic field from our full-$f$ electromagnetic turbulence simulations, the equilibrium was assumed to remain fixed. This is a decent assumption for simulations of stationary conditions in existing devices, and as our validation efforts show, the simulated pressure profiles can indeed be close to the experimental measurements \cite{zholobenko2021role, Oliveira2022, zhang2024magnetic,zholobenko2024tokamak}. 
However, in general, the background pressure profile in our simulations evolves dynamically and therefore not necessarily remains the same as $p_0$ in the fixed magnetic equilibrium \eqref{eqn:equi:balance} -- the question arises naturally whether the equilibrium magnetic field should be dynamically evolved as well. 
Additionally, we note that the force balance \eqref{eqn:vort} solved by full-$f$ turbulence codes differs (even in a time average) from the standard MHD force balance \eqref{eqn:equi:balance}: it contains additionally the Reynolds and Maxwell stress terms  \cite{scott2005energetics,zholobenko2021electric,zholobenko2024tokamak}, in the polarisation flux and the parallel current flux along the perturbed magnetic field, which modify the equilibrium force balance through turbulent fluctuations. Particularly in the plasma edge, this effect could be significant, and is unaccounted for in traditional equilibrium reconstructions. 
This poses the question whether the equilibrium could be evolved self-consistently with the turbulence?

A possible way forward might be removing the Shafranov shift from the initial magnetic equilibrium but merging the $A_s$ into the magnetic equilibrium by retracing the $A_s$-perturbed magnetic field lines dynamically.
This approach effectively separates the handling of magnetic perturbations: the low-frequency magnetic shift is addressed via retracing, while high-frequency magnetic fluctuations remain in the flutter operator.
As demonstrated in Fig.~\ref{fig:flutter_approx_single}, flutter and retracing results converge to each other in the limit of small magnetic perturbations, verifying the feasibility of retracing $A_s$.
Naturally, such a procedure comes with additional computational cost.

By evolving the equilibrium, field-aligned codes might approach the full-\textit{f} magnetic field treatment. This is most clearly useful for transient simulations, such as power ramps or confinement state transitions, as well as predictive simulations of yet non-existent fusion reactors. 
The biggest challenge would be to enable studies of large-scale MHD events such as neoclassical tearing modes \cite{pamela2017recent}, type-I ELMs \cite{cathey2020non} or vertical displacement events \cite{gruber1993vertical,clauser2019vertical} within the scope of field-aligned turbulence simulations. 
The simulations of the interaction between macroscopic MHD and microscopic turbulence could reveal the structure of magnetized plasma in fusion devices from a new perspective.


\section*{Acknowledgement}
This work has been carried out within the framework of the EUROfusion Consortium, funded by the European Union via the Euratom Research and Training Programme (Grant Agreement No. 101052200 – EUROfusion). Views and opinions expressed are however those of the author(s) only and do not necessarily reflect those of the European Union or the European Commission. Neither the European Union nor the European Commission can be held responsible for them.
This work has been granted access to the HPC resources of the EUROfusion High-Performance Computer (Marconi-Fusion)
under the Project TSVV3. 
\section*{Appendix}\label{appe}
\subsection*{Divergence-free magnetic flutter implementation}\label{sec:flutter}
The total magnetic field in our model is expressed as $\mathbf{B}_\mathrm{tot} =\mathbf{B}_{0} + \mathbf{B}_{1} $, where $\mathbf{B}_{0}$ is the equilibrium field and $\mathbf{B}_{1}$ is the disturbances.
The parallel disturbance $\mathbf{B}_{1,\parallel}$ is not taken into account, namely we ignore the high-frequency magnetic compression effect \cite{zeiler1997nonlinear}.
The reason is that the flute mode ordering $k_\perp \gg k_\parallel$ implies the frequency of the compressional wave $k_\perp v_A$ must be much faster than the shear Alfv\'en frequency $k_\parallel v_A$.
We hereby assume that the magnetic disturbances $\mathbf{B}_{1}=\mathbf{B}_{1,\perp}$ are purely perpendicular to the background magnetic field, and therefore they have the shear Alfvén form $\mathbf{B}_{1} =\nabla \times A_{1} \mathbf{b}_0 $. Here, $A_1$ is the delta-\textit{f} extraction from the parallel magnetic potential $A_\parallel$ by \eqref{eqn:evol_apar} and \eqref{eqn:evol_A_1}. $\mathbf{b}_0 = \mathbf{B}_0/B$ is the unit vector of the background field.
$B$ is the strength of the total magnetic field, approximated as $B={B}_\mathrm{tot}\approx {B}_0$ due to the fact ${B}_1\ll {B}_0$.
GRILLIX consistently restricts all perpendicular dynamics within poloidal planes \cite{stegmeir2019global}. Hence we replace $\mathbf{b}_0$ with the toroidal unit vector $\mathbf{b}_\varphi=\mathbf{B}_\varphi/{B}_\varphi$ and obtain
$\mathbf{B}_{1} =\nabla \times A_{1} \mathbf{b}_\varphi $.
The perturbation vector is $\mathbf{b}_{1} =\mathbf{B}_{1}/B \approx \mathbf{B}_{1}/B_\varphi$.

The common approach of computing $\mathbf{B}_{1}$, used in many codes \cite{xia2015nonlinear,zhu2018gdb,giacomin2022gbs,kendl2010nonlinear,michels2022full}, was 
\begin{equation}\label{eqn:B_1_divt}
\mathbf{B}_{1} =\nabla \times \mathbf{b}_\varphi A_{1} =  -\mathbf{b}_\varphi \times \nabla   A_{1}+ A_{1}\nabla \times \mathbf{b}_\varphi \approx -\mathbf{b}_\varphi \times \nabla   A_{1}\, , 
\end{equation}
in which the term $A_{1}\nabla \times \mathbf{b}_\varphi$ was dropped because  $|A_{1}\nabla \times \mathbf{b}_\varphi|/|-\mathbf{b}_\varphi \times \nabla  A_{1}| \sim \rho_s /R_0 \ll 1 $ in the drift ordering (the scale of $\mathbf{b}_\varphi$ is around the major radius $R_0$, while $A_1$ has a perpendicular scale of turbulence $\rho_s$).
This simplification also avoids the magnetic potential $A_{1}$ appearing outside the derivative operator in the calculation of $\mathbf{B}_{1}$.
Nevertheless, under this simplification $\mathbf{B}_{1}$ is not divergence-free as $\nabla \cdot (-\mathbf{b}_\varphi \times \nabla A_{1}) =-\nabla A_{1} \cdot \nabla \times \mathbf{b}_\varphi\neq 0$ (the background field $\mathbf{B}_0$ is divergence-free itself.).
Clearly, $\nabla \cdot \mathbf{B}_{1}\neq 0$ is potentially dangerous, even if the error is small. 
The flutter divergence of a parallel flux variable $\Gamma_\parallel$ under this approximation takes the form
\begin{equation}\label{eqn:flutter_div_divt}
\nabla \cdot \Gamma_\parallel \mathbf{b}_{1} =\mathbf{B}_{1}\cdot\nabla\frac{\Gamma_\parallel}{B}+\frac{\Gamma_\parallel}{B}\nabla \cdot\mathbf{B}_{1}\approx \mathbf{B}_{1}\cdot\nabla\frac{\Gamma_\parallel}{B}.
\end{equation}
In principle, the ${\Gamma_\parallel}/{B}\nabla \cdot\mathbf{B}_{1}$ term should be kept as it is non-zero.
Without it, \eqref{eqn:flutter_div_divt} is not a pure divergence, and 
the flux conservation is broken, 
\begin{equation}\label{eqn:conserv_divt}
\iiint \mathbf{B}_{1}\cdot\nabla\frac{\Gamma_\parallel}{B} \mathrm{d}V \neq 
\oiint \Gamma_\parallel \mathbf{b}_{1}\cdot \mathrm{d}\mathbf{S}\, ,
\end{equation}
where $\mathrm{d}V$ is the volume element between flux surfaces and $\mathrm{d}\mathbf{S}$ is flux surface element.

To avoid the aforementioned series of issues, it is desirable to treat the magnetic fluctuations $\mathrm{B}_1$ as divergence-free in the first place. 
Instead of splitting $\mathrm{B}_1$ according to \eqref{eqn:B_1_divt}, it is sufficient to do it in terms of $A_{1}/B_\varphi$, 
\begin{equation}\label{eqn:0div_B1}
\mathbf{B}_{1} =\nabla \times (\frac{A_1}{B_\varphi}\mathbf{B}_\varphi)=-\mathbf{B}_\varphi \times \nabla \frac{A_1}{B_\varphi}+\frac{A_1}{B_\varphi}\nabla \times \mathbf{B}_\varphi=-\mathbf{B}_\varphi \times \nabla \frac{A_{1}}{B_\varphi}\, .
\end{equation}
In the magnetic equilibrium that we construct, the toroidal field fulfills the vacuum field assumption, i.e. $\mathbf{B}_\varphi$ is exclusively induced by the external coils, implying $\nabla \times \mathbf{B}_\varphi=0$ everywhere in the plasma. 
Hence, the last reduction in \eqref{eqn:0div_B1} is mathematically precise instead of an approximation. 
$\mathbf{B}_{1}$ calculated by \eqref{eqn:0div_B1} will be precisely divergence-free. 

Following \eqref{eqn:0div_B1}, the flutter gradient of a variable $f$ takes the shape of a Possion bracket
\begin{equation}\label{eqn:flutter_grad}
\mathbf{b}_{1} \cdot \nabla f =-({\mathbf{b}_\varphi} \times \nabla \frac{A_1}{B_\varphi}) \cdot \nabla f = -[\frac{A_1}{B_\varphi},f]\, .
\end{equation}
Thanks to $\nabla \cdot \mathbf{B}_1=0$, the flutter divergence can be formulated as
\begin{equation}\label{eqn:flutter_div}
\nabla \cdot f\mathbf{b}_{1}  = \mathbf{B}_1 \cdot \nabla \frac{f}{B_\varphi} 
= -B_\varphi[\frac{A_{1}}{B_\varphi},\frac{f}{B_\varphi}]\, ,
\end{equation}
which is a pure divergence ensuring the flux conservation
\begin{equation}\label{eqn:conserv_divf}
\iiint \nabla \cdot \Gamma_\parallel \mathbf{b}_{1}  \mathrm{d}V =  
\oiint \Gamma_\parallel \mathbf{b}_{1}\cdot \mathrm{d}\mathbf{S}\,.
\end{equation}

We compared the two implementations of divergent/divergence-free magnetic fluctuations by running two H-mode simulations from $t=0$ with the filter width $\lambda=10\tau_0$ for both.
After the turbulence saturates from $t=60\tau_0$, the two cases exhibit almost identical outboard mid-plane profiles, radial heat transport and particle transport (for example, the radial heat flux across the flux surface $\rho_\mathrm{pol}=0.99$ is $2.03$MW for $\nabla \cdot \mathbf{B}_1 \neq 0 $ and $2.00$MW for $\nabla \cdot \mathbf{B}_1 = 0 $). Transiently, before saturation, somewhat larger deviations do occur. Hence, as expected, the divergence-free implementation is only slightly better than the divergent one, but it also costs only little effort.

\bibliographystyle{vancouver}
\bibliography{main.bib}

\end{document}